\documentclass[ip,twocolumn]{jpsj3}

\usepackage{bm,amsmath}
\usepackage{txfonts}
\usepackage[whole]{bxcjkjatype} 
\usepackage{graphicx}
\usepackage{xcolor}

\def\ket#1{|#1\rangle }
\def\bra#1{\langle #1 |}
\def\braket#1{\langle #1 \rangle}
\def\n{\nonumber \\ }

\title{Geometric Aspects of Nonlinear and Nonequilibrium Phenomena}

\author{Takahiro Morimoto$^1$, Sota Kitamura$^1$ and Naoto Nagaosa$^{1,2}$ }
\inst{$^1$Department of Applied Physics, The University of Tokyo, Hongo, Tokyo, 113-8656, Japan \\
$^2$RIKEN Center for Emergent Matter Science (CEMS), Wako, Saitama 351-0198, Japan.} 

\abst{
We review recent developments in the research of nonlinear and nonequilibrium phenomena in solids focusing on their geometrical aspects. We start with introducing the basic concepts of geometrical phases of Bloch electrons and Floquet theory for periodically driven systems. Then we review recent attempts to engineer topological phases in nonequilibrium matters such as graphene, magnets, and superconductors irradiated with circularly polarized light.
We next review a bulk photovoltaic effect of inversion broken materials focusing on the shift current response. The shift current is described with Berry connections and has a close relationship to the modern theory of polarization. We further review recent extensions of the shift current to correlated electron systems. Finally, we explain the geometric diabatic time evolution in the Zener tunneling process and its consequences on nonreciprocal transport.
}

\begin{document}
\maketitle

\section{Introduction}
The notion of geometry is serving as an important guiding principle in recent research on condensed matters. In particular, the discovery of quantum phases protected by symmetry and topology of wave functions in solids has boosted the search for geometrical phenomena in quantum materials, as exemplified by topological insulators \cite{hasan-kane10,qi-zhang11} and Weyl/Dirac semimetals \cite{armitage18}. The topological phases of matter support stable gapless excitations and exhibit unique response phenomena stemming from the nontrivial topology of Bloch wave functions. For example, 2D and 3D TIs support spin Hall effect \cite{kane-mele1,kane-mele2} and topological magnetoelectric effect \cite{qi-hughes-zhang,essin09}, and Weyl semimetals support transport phenomena from chiral anomaly\cite{burkov15}. Topological superconductors in 2D support Majorana fermions with its possible application to quantum computation\cite{sato17}. While the geometrical aspects of those phases and responses have been intensely studied, the focus of the studies was mainly on the quantum phases in the equilibrium and response phenomena within the linear response regime. 
Nonlinear and nonequilibrium phenomena are a fertile field of research given their potential for novel functionality, yet how the nontrivial geometry of electronic wave functions in solids affect nonlinear and nonequilibrium phenomena have remained less explored.

Nonlinear responses of electrons in bulk crystals have a long history of research and include important phenomena such as frequency conversions, nonlinear Kerr effect, and rectification effect, which have been studied actively both in fundamental science and application purposes.\cite{Boyd,Bloembergen,Sturman}
In particular, when inversion symmetry is broken in crystals, second-order nonlinear responses are allowed, including the bulk photovoltaic effect (BPVE) and nonreciprocal transport, which can be applied to solar cells and diodes. 
With the recent discovery of highly efficient BPVE in perovskite oxides, \cite{Nie,Shi,deQuilettes} the BPVE in inversion broken materials with shift current mechanism is attracted keen attention as a way to overcome the Shockley–Queisser limit for conventional solar cells based on p-n junctions. Shift current arises from the shift of electron wave packet during the optical transition and has a geometrical origin, in that the wave packet shift is described by Berry connection of the Bloch electrons and has a close relationship to the modern theory of electric polarization.\cite{Baltz,Sipe,Young-Rappe,Morimoto-Nagaosa16}
BPVE from the shift current mechanism has been studied actively including inversion broken semiconductors,\cite{Sotome19} Weyl semimetals,\cite{osterhoudt19} and transition metal dichalcogenides.\cite{Akamatsu21}
Moreover, the research of BPVE is extended to correlated electron systems such as systems with excitons,\cite{Morimoto-exciton16} magnetic systems,\cite{Morimoto-magnon19,Morimoto-magnon21} 
phonons \cite{Okamura22} and superconductors.\cite{xu19}

Another important direction to explore geometrical responses beyond the linear response regime is the nonequilibrium systems. A representative example is a nonequilibrium steady state realized in graphene under strong circularly polarized light which exhibits a quantized anomalous Hall effect.\cite{Oka2009,Kitagawa2011} The nonequilibrium systems with a periodic driving such as an external light field can be treated with the Floquet theory.\cite{Eckardt2017,Oka2019,Rudner2020,Torre2021} Floquet theory is regarded as a time direction analog of Bloch theorem that describes electrons in solids with spatially periodic potentials of ions.
Specifically, Floquet theory can describe nonequilibrium systems with an effectively static Hamiltonian (Floquet Hamiltonian) in an extended Hilbert space with the Floquet index.\cite{Shirley1965,Sambe1973} A perturbative treatment on the Floquet Hamiltonian often gives an effective Hamiltonian that exhibits qualitatively different behavior from original systems \cite{Eckardt2015,Bukov2015,Mikami2016} and allows control of quantum phases without changing the chemical composition of solids. Such engineering of exotic phases with the external field is actively studied under the name of ``Floquet engineering'' in recent years.\cite{Oka2019}
For example, irradiation of magnets with circularly polarized light is proposed to generate scalar chirality and topologically nontrivial magnetic structure.\cite{Kitamura2017,Claassen2017}  Superconductors with circularly polarized light can exhibit topological superconductivity.\cite{Kitamura2022} 
On the experimental side, the Floquet bands are observed for the Dirac fermions at the surface of topological insulators,\cite{wang2013observation} and the anomalous Hall effect in the circularly driven systems has been realized in cold atoms.\cite{Jotzu2014} 

In this paper, we review recent advances in the research of geometrical responses in nonlinear and nonequilibrium phenomena. In Sec. II, we introduce basic concepts of geometrical phases in solids including Berry connection and Berry curvature. In Sec. III, we introduce Floquet theory for periodically driven systems and explain basic notions including Sambe space representations and the high frequency expansion in deriving effective Hamiltonians. 
In Sec. IV, we review salient examples of Floquet engineering starting from the graphene with circular driving and extensions in strongly correlated systems including magnetic systems and superconductors. In Sec. V, we review geometrical aspects of bulk photovoltaic effects focusing on the shift current and its recent developments in correlated electron systems. 
In Sec. VI, we review geometric phenomena in diabatic time evolution in the tunneling process exhibiting a nonperturbative response. In Sec. VII, we give a summary of the review article and discuss future directions.

\section{Geometry in Solids \label{sec: berry}}
In this section, we review the basics of the geometrical properties of Bloch electrons in solids (Fig.~\ref{fig: berry connection}). 

\subsection{Berry connection as an intracell coordinate}
\begin{figure}
    \centering
    \includegraphics[width=\linewidth]{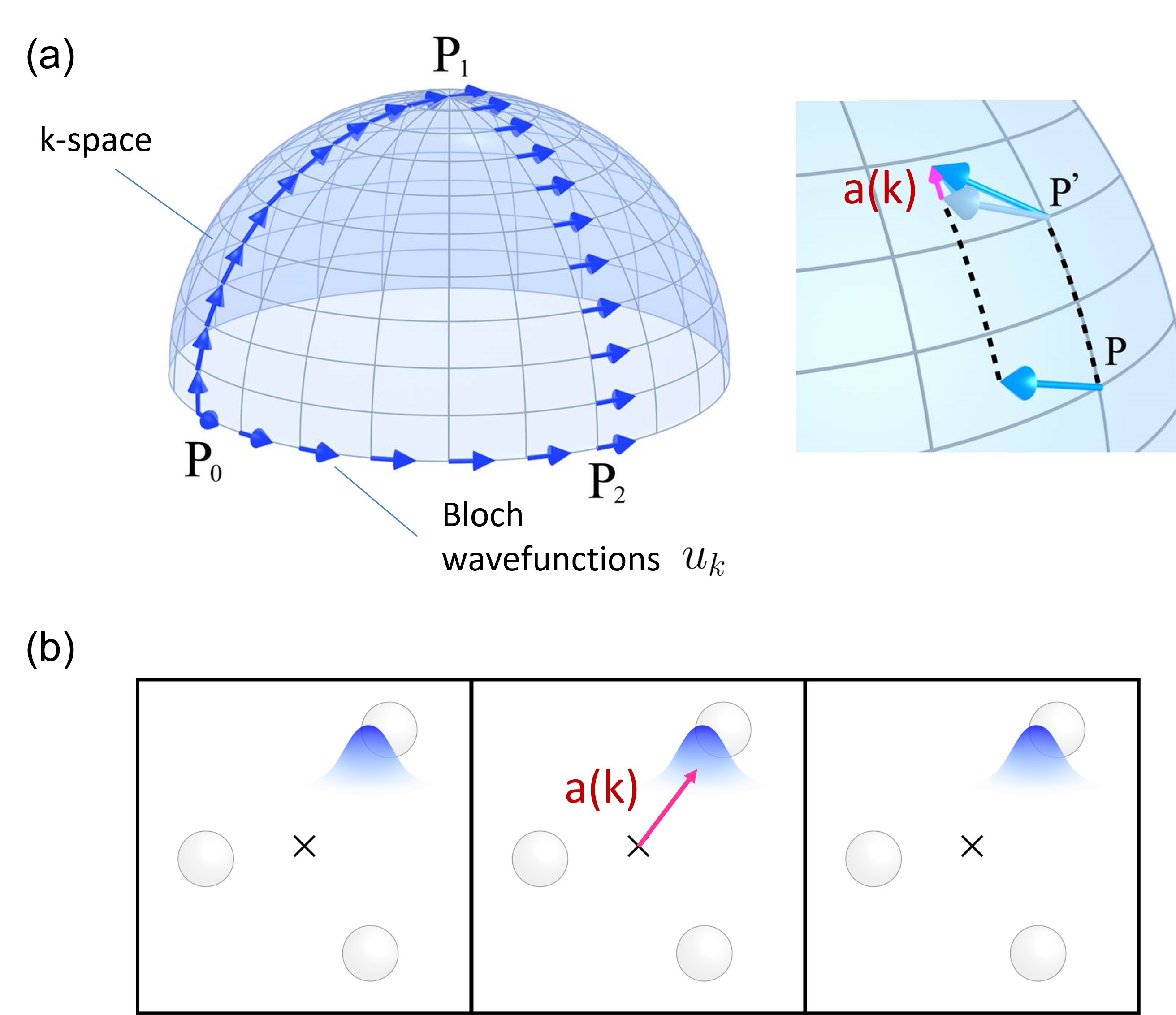}
    \caption{(Color Online)
    Geometry of Bloch electrons in solids.
    (a) The momentum space picture of Berry connection. 
    Bloch wave functions $u_k$ are represented as arrows defined at each point in the momentum space. The Brillouin zone forms a curved space for a given band.
    The parallel transport of $u_k$ in the momentum space gives rise to a nontrivial phase that is quantified by the Berry connection $a(k)$. 
    Adapted from Ref.~\citen{Nagaosa-Morimoto17}. \copyright[2017]{Wiley.}
    (b) The real space picture of Berry connection. Black squares represent unit cells with ions (white circles) inside. The wave packet made from the Bloch states (indicated by blue) is generally located away from the center of the unit cell. The Berry connection $a(k)$ measures this intracell coordinate of the Bloch state.
    }
    \label{fig: berry connection}
\end{figure}

Let us consider an electron in a periodic potential $V(r)$, which is described by a Schr\"{o}dinger equation,
\begin{align}
    H\psi&=E \psi, &  
    H&= -\frac{\hbar^2}{2m} \nabla^2 + V(r),
\end{align}
Here, the periodic potential satisfies $V(r+a)=V(r)$ with the lattice constant $a$.
According to Bloch's theorem, the wave function for an eigenstate can be written as 
\begin{align}
    \psi_{nk}(r)=
    u_{nk}(r)e^{ikr},
\end{align}
where $n$ is the band index, $k$ is the momentum, 
and $u_{nk}$ is the periodic part of the wave function satisfying $u_{nk}(r+a)=u_{nk}(r)$.
The periodic part of the wave function $u_{nk}$ and the band dispersion $E_n(k)$ satisfy
\begin{align}
H(k) u_{nk}(r) = E_n(k) u_{nk}(r),
\end{align}
where $H(k)$ is the Bloch Hamiltonian obtained as a Fourier transform of $H$.
Next, let us consider the position operator for the Bloch electrons.
Since the Bloch wave function is extended over the system, we consider a wave packet so that the position expectation value is well-defined.
Specifically, we define a Bloch wave packet by
\begin{align}
    \Psi_B(r) = 
    V_0 \int [dk] c(k) \psi_{nk}(r),
\end{align}
where $V_0$ is the volume of the unit cell, and $c(k)$ is some function that has nonzero values only around some momentum $k=k_0$.
Here we define $[dk]\equiv dk/(2\pi)^d$ for a $d$ dimensional system.
The position expectation value $\braket{r}$ for $\Psi_B$ can be computed as
\begin{align}
    \braket{r} &= \int dr \Psi_B^*(r) r \Psi_B(r) \n
    &= V_0^2 \int [dk][dk'] dr
    c^*(k')  u_{nk'}^*(r) u_{nk}(r) c(k) r e^{i(k-k')r} \n
    &=V_0^2 \int [dk][dk'] dr
    c^*(k')  u_{nk'}^*(r) u_{nk}(r) c(k) (-i\partial_k) e^{i(k-k')r} \n
    &= V_0^2 \int [dk][dk'] dr
    c^*(k') i\partial_k [ u_{nk'}^*(r) u_{nk}(r) c(k)] e^{i(k-k')r} \n 
    &= V_0 \int [dk]dk' c^*(k') i\partial_k  [\braket{u_{nk'} | u_{nk}} c(k)] \delta(k-k') \n
    &= V_0 \int [dk]  c^*(k) [ i \partial_k  + i   
    \langle u_{nk} |\partial_k  u_{nk} \rangle ] c(k).
    \label{eq: r covariant derivative}
\end{align}
When moving into the fourth line, we decomposed the integration over $r$ into that over the unit cell and a summation over the unit cells,
where the former leads to $\braket{u_{nk'} | u_{nk}}$ and the latter $((2\pi)^d/V_0)\delta(k-k')$.
(We adopted a convention that $\psi_{nk}$ is normalized to one unit cell, $\int_{V_0} dr|\psi_{nk}(r)|^2=1$~\cite{marzari-rmp}.)
The above equation indicates that an effective position operator $r$ for the Bloch wave packets constructed from the $n$th band is given by
\begin{align}
    r=i\partial_k + a_n(k),
\end{align}
with 
\begin{align}
    a_n(k)=i\braket{u_{nk}|\partial_k u_{nk}}.
\label{eq:Berry}
\end{align}
Here, $a_n(k)$ is the Berry connection and describes the position of the Bloch wave packet within the unit cell (i.e. the intracell coordinate).\cite{Resta}
While it is expected that the position operator in the momentum space representation is given by $k$-derivative from the canonical conjugate relationship between $r$ and $k$, $i\partial_k$ itself depends on the gauge choice of the Bloch wave functions.
This additional term $a_n$ arises from the fact that an electron is energetically constrained within a particular Bloch band and the combination $i\partial_k + a_n(k)$ becomes gauge invariant, which is known as a covariant derivative.
The geometric notions such as Berry connection and the covariant derivative naturally appear in the expression for physical quantities because the wave function of Bloch electrons of a particular band forms a curved space, i.e., the Bloch wave function is $k$ dependent in contrast to the full Hilbert space is $k$ independent over the Brillouin zone (Fig~\ref{fig: berry connection}).
We note that this description of the position operator is valid as far as one focuses on the low energy dynamics of a Bloch electron in a particular band and transition into states in the other bands can be neglected.
From the Berry connection, we define the Berry curvature as
\begin{align}
    F_n(k) = \bm \nabla \times \bm{a_n}(k),
\end{align}
which quantifies the nontrivial curvature of the wave function for the band $n$ over the momentum space.

Here, a few comments are in order.
First, we note that the Berry phase and Berry curvature are defined for the wave function 
confined within an eigenstate separated from others energetically,
while those for the Bloch wave function are defined for the periodic part as 
given in Eq.~(\ref{eq:Berry}). 
The reason is clear from Eq.~(\ref{eq: r covariant derivative}) that the 
connection  $a_n(k)$ represents the intracell coordinate in addition to the $k$ derivative.\cite{vanderbilt_2018} 
In practice, the eigenstate $\ket{u_{nk}}$ is obtained by diagonalizing the Hamiltonian 
matrix, where the basis for this matrix should be carefully chosen. 
In the orbital basis, one can choose the basis as
$\Phi_{\alpha}(r) = \sum_{R_n} \phi_\alpha(r-R_n)$ where $R_n$ is the 
center of the $n$-th unit cell and $\alpha$ specifies the atom and orbital in each
unit cell. The tight-binding model uses this basis, which is independent of
the momentum $k$. 
A slightly different choice of the orbital-basis is 
$\Phi'_{\alpha k}(r) = \sum_{R_n} e^{-ik(r-R_n)} \phi_\alpha(r-R_n)$
which depends on $k$. 
Another basis is used in the $k \cdot p$-theory. In that case, the basis is chosen as
$u_{n}(r) = u_{n k_0}(r)$
which is the eigenstate at the reference momentum $k_0$ and again has no $k$ dependence. The Hamiltonian is diagonal at
$k_0$ and is given by expanding with respect to $k-k_0$.
Lastly, the Hamiltonian is diagonal in the basis of the Bloch wave function 
$u_{nk}(r)$ by definition. When the additional 
Hamiltonian is added such as the 
coupling to the electromagnetic field,
one needs to take into account the
$k$ dependence of $u_{nk}(r)$ for the 
diagonalization. The dependence of the basis 
function requires some care since $k$ 
derivative of the basis function should be 
taken into account in that case.
For example, the matrix 
elements of $r$ between the different bands 
$n$ and $m$ is given by
\begin{equation}
    \langle \psi_{nk} | r | \psi_{mk} \rangle 
    = a_{nm}(k)
\end{equation}
where 
\begin{equation}
     a_{nm}(k) = i \langle u_{nk} | \partial_k | u_{mk} \rangle
\end{equation}
is the off-diagonal element of the non-Abelian Berry connection. This is the fundamental equation for the calculation in the length gauge.

\subsection{Berry curvature and anomalous velocity}
The nontrivial geometry of the Bloch electrons appears in the transport phenomena.
Let us consider the motion of a Bloch wave packet in the presence of a dc electric field.
We assume the electric field is applied along the $x$ direction and consider the Hamiltonian $H-qEx$, where $q$ denotes the charge of an electron.
(Here, we use the notation $q=-|e|$ for electron charge.)
The velocity of the Bloch electron is obtained with the Heisenberg equation as
\begin{align}
    \braket{v}=-\frac{i}{\hbar} \braket{[r,H(k)-qEx]}.
\end{align}
Focusing on the low energy dynamics of electrons in the band $n$, we use the position operator in the covariant derivative form $r=i\partial_{k}+a_n(k)$.
The $x$ component of the velocity is given by
\begin{align}
    \braket{v_x}=\frac{1}{\hbar} \braket{\partial_{k_x}H(k)}
    =\frac{1}{\hbar}\partial_{k_x}E_n(k),
\end{align}
which is the group velocity along the $x$ direction.
Similarly, the $y$ component of the velocity is given by
\begin{align}
    \braket{v_y}
    &=-\frac{i}{\hbar}\braket{[y,H(k)]}
    -\frac{iq}{\hbar}E\braket{[x,y]}
\end{align}
The first term again gives the group velocity along the $y$ direction. 
The second term contains the commutator of $x$ and $y$, which can be expressed with the Berry curvature as
\begin{align}
    [x,y] &= [i\partial_{k_x} + a_{x,n}, i\partial_{k_y} + a_{y,n}]
=i \partial_{k_x} a_{y,n} - i\partial_{k_y} a_{x,n}
= i F_{z,n}(k).
\end{align}
This equation indicates that the Berry curvature measures the noncommutativity of the $x$ and $y$ directions for Bloch electrons in solids.
Thus we obtain
\begin{align}
\braket{v_y}=  \frac{1}{\hbar}\partial_{k_y} E_n(k) + \frac{q}{\hbar} E F_n(k), 
\end{align}
where the second term is called the anomalous velocity and arises from the nontrivial geometry of the momentum space.
Anomalous velocity is the transverse (Hall) current response to the applied electric field and is the origin of the anomalous Hall effect in materials with broken time reversal symmetry such as magnetic materials \cite{nagaosa-rmp}.

When a band is fully occupied in insulators, its contribution to the Hall current is solely expressed with the Berry curvature as
\begin{align}
J_y = q \int [dk] \braket{v_y} = \frac{q^2}{\hbar} E \int [dk] F(k),
\end{align}
since the contribution from the group velocity cancels in the integral.
The Hall conductivity $\sigma_{xy}$ is obtained from the relations, $J_y = \sigma_{yx} E_x$ and $\sigma_{xy}=-\sigma_{yx}$,
as
\begin{align}
\sigma_{xy} = - \frac{q^2}{\hbar} \int [dk] F(k).
\label{eq: sigma xy}
\end{align}
In particular, the Hall conductivity in two dimensions can be expressed as
\begin{align}
\sigma_{xy} = - \frac{q^2}{h} \int \frac{d^2k}{2\pi} F(k)= - \frac{q^2}{h} C,
\end{align}
where $C$ is the Chern number that takes an integer value.
This gives a semiclassical picture of the quantum Hall effect.

\subsection{Berry phase and electric polarization}
Berry connection has a meaning of the intracell coordinate and is closely connected with the electric polarization induced by the positional shift of the Bloch electrons within the unit cell, which is known as ``the modern theory of electric polarization''~\cite{Resta,Kingsmith93,Vanderbilt93}.
Due to the periodic structure of crystals, it is necessary to specify the base point for measuring the positional shift of electrons. 
To this end, we introduce the Wannier function which is spatially localized at some unit cell and is written with Bloch wave functions as~\cite{marzari-rmp} 
\begin{align}
W_{n,R}(r)= 
V_0
\int [dk]
e^{-ikR}\psi_{n,k}(r),
\end{align}
where $R$ is the lattice vector specifying the unit cell, and $V$ is the volume of the system.
The electric polarization is given by the position expectation value of the Wannier function for the unit cell $R=0$, which leads to
\begin{align}
P_n = q \int dr  W_{n,R=0}^*(r) r W_{n,R=0}(r).
\label{eq: Pn with W}
\end{align}
We can evaluate this by setting $c(k)=1$ in Eq.~\eqref{eq: r covariant derivative}
and obtain
\begin{align}
P_n = q V_0 \int [dk] a_n(k).
\end{align}
In particular, in one dimension, this can be written with the Berry phase $\theta_n=\int dk a_n(k)$ as
\begin{align}
P_n= \frac{qa}{2\pi} \theta_n,
\end{align}
with the lattice constant $a$,
which is known as the Berry phase formula for the electric polarization.~\cite{Resta,Vanderbilt93}
Berry phase is invariant under a local gauge transformation, which means $P_n$ is well described in the $k$ space representation.
Meanwhile, a large gauge transformation 
$u(r) \to u(r)e^{-ikR}$ leads to a change of the Berry phase $\theta \to \theta + 2\pi R/a$.
$R/a$ is an integer and corresponds to the choice of the base point (or the unit cell) for measuring the electric polarization.
This essentially has the same meaning as using the Wannier function $W_{n,R}$ localized at a different unit cell instead of $W_{n,R=0}$ in evaluating $P_n$ in Eq.~\eqref{eq: Pn with W},
which naturally leads to the shift of the electric polarization by $qR/a$ from the $i\partial_k$ term in Eq.~\eqref{eq: r covariant derivative}.

\section{Floquet Theory}
In this section, we review the Floquet theory, a theoretical method to handle periodically-driven nonequilibrium problems beyond the usual perturbative formulation.\cite{Eckardt2017,Oka2019,Rudner2020,Torre2021}
We introduce the Floquet theory from the viewpoint of discrete translational symmetry on time,
where the corresponding conservation law leads to the concept of effective static Hamiltonian.

\subsection{Discrete time translation and its eigenstates}

Periodically-driven systems are characterized by a time-dependent Hamiltonian $\hat{H}(t)$ with a periodicity
\begin{equation}
\hat{H}(t)=\hat{H}(t+T),
\end{equation}
where $T$ is the driving period. The Hamiltonian governs the time evolution of the wave function via the time-dependent Schr\"odinger equation
\begin{equation}
i\hbar\dfrac{\partial}{\partial t}|\Psi(t)\rangle=\hat{H}(t)|\Psi(t)\rangle,\label{eq:tdse}
\end{equation}
whose solution can be formally written using the time-evolution operator,
\begin{align}
\hat{U}(t,t_{0})  & =\sum_{N=0}^{\infty}\dfrac{(-i)^{N}}{\hbar^N N!}\int_{t_{0}}^{t}dt_{1}\dots\int_{t_{0}}^{t}dt_{N}\mathcal{T}[\hat{H}(t_{1})\dots\hat{H}(t_{N})]\notag\\
& =\mathcal{T}\exp\left[-\frac{i}{\hbar}\int_{t_{0}}^{t}dt^{\prime}\hat{H}(t^{\prime})\right]
\end{align}
with $\mathcal{T}$ denoting the time-ordered product, as $|\Psi(t)\rangle=\hat{U}(t,t_{0})|\Psi(t_{0})\rangle$.

Typically, it is impossible to calculate $\hat{U}$ for a time-dependent problem in an analytical manner, so that we usually perform a perturbative expansion with respect to the field strength starting from an equilibrium state. While analytic calculations remain difficult even for the periodically-driven systems, we can find special properties derived from symmetry. In the symmetry viewpoint, the time periodicity can be rephrased as the invariance under the discrete time translation, and there should be an associated ``conservation law". 

\begin{figure}
    \centering
    \includegraphics[width=\linewidth]{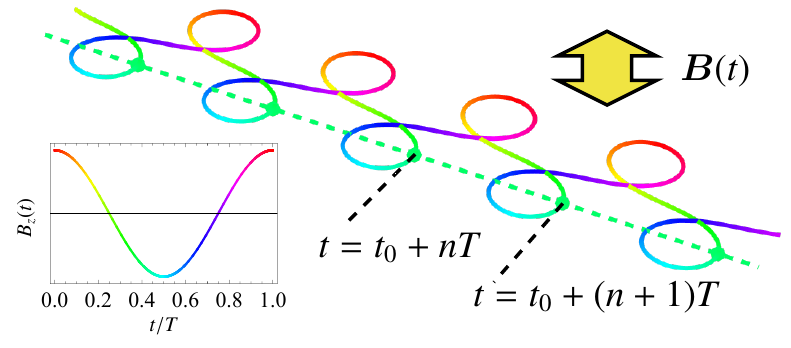}
    \caption{(Color Online)
    Schematic picture of time evolution in a periodically-driven system.
    Here we show an example of a real-space dynamics of an electron in an oscillating magnetic field.\cite{Oka2016}
    The solid curve shows the actual trajectory of the center of the wave packet by the time-periodic Hamiltonian $\hat{H}(t)$ with a period $T$, colored by the phase of the external magnetic field shown in the inset.
    The dashed line shows the effective dynamics generated by a static operator $\hat{F}$, 
    which describes a free particle motion with a renormalized effective mass.
    These two dynamics coincide at stroboscopic times $t=t_0+nT$ with $n$ being an integer.
    }
    \label{fig:stroboscopic}
\end{figure}

Here, let us formulate this ``conservation law" using the discrete time translation described by $\hat{U}(t_{0}+nT,t_{0})$ with an integer $n$.
The periodicity of the Hamiltonian leads to an invariance of $\hat{U}$ under the shift of the interval of the time integration, i.e.,
\begin{equation}
\hat{U}(t+T,t^{\prime}+T)=\hat{U}(t,t^{\prime}).
\end{equation}
Combining this with a basic property $\hat{U}(t,t^{\prime})=\hat{U}(t,t^{\prime\prime})\hat{U}(t^{\prime\prime},t^{\prime})$, we obtain 
\begin{align}
\hat{U}(t_{0}+nT,t_{0}) & =\prod_{m=1}^{n}\hat{U}(t_{0}+(n-m+1)T,t_{0}+(n-m)T)\notag\\
 & =[\hat{U}(t_{0}+T,t_{0})]^{n}.
\end{align}
Namely, the solution of Eq.~(\ref{eq:tdse}) at $t=t_0+nT$ can be written as 
\begin{equation}
|\Psi(t_{0}+nT)\rangle  =e^{-i\hat{F}nT/\hbar}|\Psi(t_{0})\rangle,
\end{equation}
where 
\begin{equation}
\hat{F}=\frac{i\hbar}{T}\ln\hat{U}(t_{0}+T,t_{0}).\label{eq:floquet-hamiltonian}
\end{equation}
This relation implies that the time evolution of a periodically-driven system is equivalent to that with a time-independent Hamiltonian $\hat{F}$ if we focus on ``stroboscopic" times $t=t_0+nT$.\cite{Shirley1965,Kitagawa2011,Bukov2015}
We show a schematic picture of this stroboscopic time evolution in Fig.~\ref{fig:stroboscopic}.

While the effective static Hamiltonian $\hat{F}$ depends on the reference time $t_0$ (modulo $T$), its eigenvalue, called quasienergy, is independent of $t_0$ as we see below.
This implies that the quasienergy is the conserved quantity associated with the discrete time translation. Let us impose an initial condition to the time-dependent Schr\"odinger equation at $t=t_0$ as 
\begin{equation}
\hat{F}|\Psi_{\alpha}(t_{0})\rangle=\epsilon_{\alpha}|\Psi_{\alpha}(t_{0})\rangle.
\end{equation}
Then we can easily show that
\begin{align}
|\Psi_{\alpha}(t+T)\rangle & =\hat{U}(t+T,t_{0}+T)\hat{U}(t_{0}+T,t_{0})|\Psi_{\alpha}(t_{0})\rangle\notag\\
 & =\hat{U}(t,t_{0})e^{-i\epsilon_{\alpha}T/\hbar}|\Psi_{\alpha}(t_{0})\rangle\notag\\
 & =e^{-i\epsilon_{\alpha}T/\hbar}|\Psi_{\alpha}(t)\rangle,
\end{align}
which reveals that the eigenstate of $\hat{F}$, which we call a Floquet state, has a discrete translation invariance at any time $t$. In particular, if we introduce $|v_\alpha(t)\rangle$ by
\begin{equation}
|\Psi_{\alpha}(t)\rangle=|v_{\alpha}(t)\rangle e^{-i\epsilon_{\alpha}t/\hbar},\label{eq:floquet-decomp}
\end{equation}
we can rewrite the above property in an analogous form to the Bloch theorem for spatially-periodic problems, as
\begin{align}
|v_{\alpha}(t+T)\rangle & =|v_{\alpha}(t)\rangle.
\end{align}
Combining this with the fact that the eigenstates of $\hat{F}$ form a complete set (as in a usual static Hamiltonian), we can write the solution of the time-dependent Schr\"odinger equation with an arbitrary initial condition as a linear combination of the Floquet states. This is known as the Floquet theorem and governs the dynamics of time-periodic systems.

Let us remark on the indefiniteness of the quasienergy $\epsilon_{\alpha}$. As in the crystal momentum for the Bloch states, the quasienergy can be distinguished only within the ``Brillouin zone" $\epsilon_\alpha\in(-\hbar\Omega/2,\hbar\Omega/2]$, where $\Omega=2\pi/T$ is the driving frequency. This is consistent with the fact that $e^{-i\Omega t}$ is a time-periodic function with a period $T$, so that the decomposition Eq.~(\ref{eq:floquet-decomp}) is not unique. It is also consistent with the fact that $\hat{F}$ defined as Eq.~(\ref{eq:floquet-hamiltonian}) cannot be uniquely determined due to the indefiniteness of the complex logarithm.

\subsection{Sambe space representation}

While we have revealed that the Floquet states govern the dynamics of periodically-driven systems, they are obtained as the eigenstates of the effective static Hamiltonian $\hat{F}$, which is still hard to calculate since it is defined from the time-evolution operator.
Here we introduce an alternative way to characterize the Floquet states without computing $\hat{F}$.

Because the Hamiltonian $\hat{H}(t)$ and the periodic part of the Floquet state $|v_\alpha(t)\rangle$ are a time-periodic function, we can expand them into Fourier series as
\begin{equation}
|v_{\alpha}(t)\rangle  =\sum_{m=-\infty}^{\infty}|v_{\alpha,m}\rangle e^{-im\Omega t},\quad
\hat{H}(t)  =\sum_{m=-\infty}^{\infty}\hat{H}_{m}e^{-im\Omega t}.
\end{equation}
Since the Floquet state is the solution of Eq.~(\ref{eq:tdse}), $i\hbar\partial_{t}(|v_{\alpha}(t)\rangle e^{-i\epsilon_{\alpha}t/\hbar})=\hat{H}(t)|v_{\alpha}(t)\rangle e^{-i\epsilon_{\alpha}t/\hbar}$ should hold, which can be rewritten as
\begin{equation}
\sum_{m^\prime=-\infty}^{\infty}[\hat{H}_{m,m^\prime}-\delta_{m,m^\prime}m\hbar\Omega]|v_{\alpha,m^\prime}\rangle=\epsilon_{\alpha}|v_{\alpha,m}\rangle,\label{eq:sambe}
\end{equation}
where we have introduced a notation $\hat{H}_{m,m^\prime}=\hat{H}_{m-m^\prime}$.
As the summation here has a form of matrix multiplication, this equation can be seen as a static Schr\"odinger equation defined on an extended Hilbert space (called Sambe space) with an additional degree of freedom labeled by the Fourier index $m$,\cite{Shirley1965,Sambe1973,Eckardt2015}
\begin{equation}
\begin{pmatrix}\ddots & \ddots & \ddots\\
\ddots & \hat{H}_{0}+\hbar\Omega & \hat{H}_{-1} & \hat{H}_{-2}\\
\ddots & \hat{H}_{+1} & \hat{H}_{0} & \hat{H}_{-1} & \ddots\\
 & \hat{H}_{+2} & \hat{H}_{+1} & \hat{H}_{0}-\hbar\Omega & \ddots\\
 &  & \ddots & \ddots & \ddots
\end{pmatrix}\begin{pmatrix}\vdots\\
\vphantom{\vdots}|v_{\alpha,-1}\rangle\\
\vphantom{\vdots}|v_{\alpha,0}\rangle\\
\vphantom{\vdots}|v_{\alpha,+1}\rangle\\
\vdots
\end{pmatrix}=\epsilon_\alpha
\begin{pmatrix}\vdots\\
\vphantom{\vdots}|v_{\alpha,-1}\rangle\\
\vphantom{\vdots}|v_{\alpha,0}\rangle\\
\vphantom{\vdots}|v_{\alpha,+1}\rangle\\
\vdots
\end{pmatrix}.
\end{equation}
Namely, the Floquet state and the quasienergy can be calculated as an eigenstate and an eigenenergy in the Sambe space, respectively. 

When we adopt a monochromatic laser light as an external driving, the Fourier index $m$ can be interpreted as the number of absorbed photons. Thus we often refer to $m$ as the photon number. 
If we quantize the photon in this setup, the Sambe space representation and the original representation with the time-dependent Schr\"odinger equation correspond to the Schr\"odinger picture and the interaction picture, respectively.~\cite{footnotesk1}

\subsection{Floquet theorem in terms of operators}
We here briefly introduce statements equivalent to the Floquet theorem in terms of operators rather than states. 
To this end, let us first write the time-evolution operator using the Floquet states as 
\begin{equation}
\hat{U}(t,t_{0})  =\sum_{\alpha}|\Psi_{\alpha}(t)\rangle\langle\Psi_{\alpha}(t_{0})|
  =\sum_{\alpha}|v_{\alpha}(t)\rangle\langle v_{\alpha}(t_{0})|e^{-i\epsilon_{\alpha}(t-t_{0})/\hbar}.\label{eq:tevol-floquet}
\end{equation}
Introducing a time-independent operator $\hat{H}_{\text{eff}}=\sum_{\alpha}\epsilon_{\alpha}|\phi_{\alpha}\rangle\langle\phi_{\alpha}|$ with an arbitrary basis set $\{|\phi_\alpha\rangle\}$, we can rewrite the time-evolution operator in a decomposed form\cite{Shirley1965,Goldman2014,Eckardt2015}
\begin{align}
\hat{U}(t,t_{0}) & =\hat{V}(t)e^{-i\hat{H}_{\text{eff}}(t-t_{0})/\hbar}\hat{V}^{\dagger}(t_{0})\label{eq:tevol-decomp}
\end{align}
with a time-periodic unitary $\hat{V}(t)=\hat{V}(t+T)=\sum_{\alpha}|v_{\alpha}(t)\rangle\langle\phi_{\alpha}|$.
In particular, we obtain $\hat{H}_{\text{eff}}=\hat{F}$ when we adopt  $|\phi_{\alpha}\rangle=|v_{\alpha}(t_{0})\rangle$.
Using the basic properties of the time-evolution operator, we can further show that $\hat{V}(t)$ satisfies a first-order differential equation \cite{Bukov2015,Mikami2016}
\begin{equation}
\hat{H}_{\text{eff}}=\hat{V}^{\dagger}(t)\hat{H}(t)\hat{V}(t)-i\hbar\hat{V}^{\dagger}(t)\frac{\partial}{\partial t}\hat{V}(t)\label{eq:heff-eom},
\end{equation}
which implies that the time-periodic unitary $\hat{V}(t)$ makes the Hamiltonian time-independent. The existence of $\hat{V}(t)$ for Eqs.~(\ref{eq:tevol-decomp}) and (\ref{eq:heff-eom}) is respectively equivalent to the Floquet theorem.

\subsection{High-frequency expansion}
The effective static problem defined on the Sambe space is quite useful as we can apply various theoretical tools developed for equilibrium systems in a straightforward manner.
The drawback of the Sambe space formalism (in a numerical viewpoint) is the limited system size due to the extension of the Hilbert space, which however can be circumvented with a perturbative treatment when the driving frequency $\Omega$ is large enough.\cite{Casas2001,Mananga2011,Goldman2014,Bukov2015,Eckardt2015,Mikami2016}
Namely, by handling the interaction changing the photon number $m$ as a perturbation, 
we can obtain an effective Hamiltonian that unchanges the photon number and can reduce the Hilbert space.
We note that here the small parameter of the series expansion is $1/\Omega$ (high-frequency expansion), and thus the formalism is still applicable to arbitrary field strength. Due to this advantage, we can employ analytical approaches even for nonequilibrium phenomena that are difficult to capture in the conventional perturbation theory. 

The high-frequency expansion can be straightforwardly obtained by applying the quasi-degenerate perturbation theory to the time-independent Schr\"odinger equation in the Sambe space Eq.~(\ref{eq:sambe}), with regarding the $-\delta_{m,m^\prime}m\hbar\Omega$ term as the unperturbed Hamiltonian.
If we adopt the formulation with the canonical transformation, this corresponds to an order-by-order determination of $\hat{V}(t)$ and $\hat{H}_\text{eff}$ in Eq.~(\ref{eq:heff-eom}), with writing $\hat{V}(t)=e^{-i\hat{\Lambda}(t)}$ and expanding Fourier components of $\hat{\Lambda}(t)=\sum_{m=-\infty}^{\infty}\hat{\Lambda}_{m}e^{-im\Omega t}$ into $1/\Omega$ series.
Up to the leading and subleading order, we obtain\cite{Bukov2015,Eckardt2015,Mikami2016}
\begin{align}
i\hat{\Lambda}^{(1)}_{m\neq0}&=-\frac{\hat{H}_{m}}{m\hbar\Omega},\\
i\hat{\Lambda}^{(2)}_{m\neq0}&=
\sum_{n\neq0}\frac{[\hat{H}_{n},\hat{H}_{m-n}]}{2mn\hbar^2\Omega^{2}}+\frac{[\hat{H}_{m},\hat{H}_{0}]}{2m^{2}\hbar^2\Omega^{2}}-\dfrac{[i\hat{\Lambda}_{0}^{(1)},\hat{H}_{m}]}{2m\hbar\Omega},\\
\hat{H}_{\text{eff}}^{(0)}&=\hat{H}_{0},\label{eq:hfe-1st}\\
\hat{H}_{\text{eff}}^{(1)}&=\sum_{m\neq0}\frac{[\hat{H}_{-m},\hat{H}_{m}]}{2m\hbar\Omega}+[i\hat{\Lambda}_{0}^{(1)},\hat{H}_{0}],\label{eq:hfe-2nd}
\end{align}
where the superscripts denote the order in $1/\Omega$.
Note that we cannot uniquely determine $\hat{\Lambda}^{(n)}_0$ just by Eq.~(\ref{eq:heff-eom}), which corresponds to the constant of integration for a first-order differential equation, $\hat{H}_{\text{eff}}=e^{i\hat{\Lambda}(t)}(\hat{H}(t)-i\hbar\partial_t)e^{-i\hat{\Lambda}(t)}$. This is related to the fact that we can choose the basis of the effective Hamiltonian arbitrarily [See Eq.~(\ref{eq:tevol-decomp})].
While the form of the effective Hamiltonian $\hat{H}_\text{eff}$ and its eigenvector $|\phi_\alpha\rangle$ depend on the choice of $\hat{\Lambda}_0$, the quasienergy $\epsilon_\alpha$ and the reconstructed Floquet state $|\Psi_\alpha(t)\rangle=e^{-i\hat{\Lambda}(t)}|\phi_\alpha\rangle e^{-i\epsilon_\alpha t/\hbar}$ does not depend on $\hat{\Lambda}_0$ within the truncation error.

The same indefiniteness of $\hat{\Lambda}$ also arises in applications of the canonical transformation to the equilibrium problems, and usually $\hat{\Lambda}^{(n)}_0=0$ is chosen, which corresponds to van Vleck's quasidegenerate perturbation theory. \cite{Bukov2015,Eckardt2015}
On the other hand, for periodically-driven systems, we sometimes adopt $\hat{\Lambda}^{(n)}(t_0)=0$ [i.e. $\hat{\Lambda}_{0}^{(n)}=-\sum_{m\neq0}\hat{\Lambda}_{m}^{(n)}e^{-im\Omega t_0}$] as well, 
since in this case we obtain $\hat{H}_\text{eff}=\hat{F}=i\hbar T^{-1}\ln\hat{U}(t_{0}+T,t_{0})$, which is directly related to time evolution. 
Note that the effective Hamiltonian depends on the reference time $t_0$ in this case.
This expansion is called Floquet-Magnus expansion.\cite{Casas2001,Mananga2011}

Let us comment on the calculation of expectation values under the high-frequency expansion.
An expectation value of a static operator $\hat{O}$ for the Floquet state is given as $\langle\Psi_\alpha(t)|\hat{O}|\Psi_\alpha(t)\rangle=\langle\phi_{\alpha}|e^{i\hat{\Lambda}(t)}\hat{O}e^{-i\hat{\Lambda}(t)}|\phi_{\alpha}\rangle$,
while we often adopt $\langle\phi_{\alpha}|\hat{O}|\phi_{\alpha}\rangle$ alternatively, nevertheless it differs from the correct expression in general.
For the Floquet-Magnus expansion, the boundary condition $\hat{\Lambda}(t_0)=0$ leads to $\langle\Psi_\alpha(t_0)|\hat{O}|\Psi_\alpha(t_0)\rangle=\langle\phi_{\alpha}|\hat{O}|\phi_{\alpha}\rangle$, 
so that the alternative expression represents the expectation value at the stroboscopic time $t=t_0+nT$.
On the other hand, we are often interested in the time average rather than the value at a specific time, 
which can be expanded as
\begin{align}
\int_0^T\frac{dt}{T}\langle\Psi_\alpha(t)|\hat{O}|\Psi_\alpha(t)\rangle & =\langle\phi_{\alpha}|(\hat{O}+[i\hat{\Lambda}_{0}^{(1)},\hat{O}])|\phi_{\alpha}\rangle+O(\Omega^{-2}),
\end{align}
which implies that the van Vleck expansion with 
$\hat{\Lambda}_{0}^{(1)}=0$
generically yields the smallest error between 
the alternative expression $\langle\phi_{\alpha}|\hat{O}|\phi_{\alpha}\rangle$
and the time average.

\subsection{Thermalization and prethermalization}

So far, we have discussed that the time evolution of periodically-driven systems can be mapped to that of a static problem. 
Here let us discuss how statistical properties of the periodically-driven systems can also be characterized by the effective static Hamiltonian, which is however quite a nontrivial problem and includes various open questions.
We can immediately notice that the straightforward extension of the equilibrium statistical mechanics in terms of the effective Hamiltonian fails, because if we naively write down the Gibbs ensemble 
$\hat{\rho}=e^{-\beta\hat{F}}/\text{Tr}e^{-\beta\hat{F}}$, it turns out to be ill-defined due to the indefiniteness of the quasienergy modulo $\hbar\Omega$.

On the other hand, the recent developments in statistical physics reveal that long-time evolution under a static nonintegrable Hamiltonian generically results in thermalization toward equilibrium states even in the absence of heat baths.\cite{Polkovnikov2011,Mori2018,Ueda2020} The formulation of this thermalization phenomenon enables us to bridge temporal and statistical properties, and indeed it is also useful for periodically-driven systems.

Thermalization of an isolated system with a static Hamiltonian can be explained using two key concepts: the diagonal ensemble and the eigenstate thermalization.\cite{Deutsch1991,Srednicki1994,Rigol2008,Reimann2008}
Let us first introduce the relaxation in terms of the diagonal ensemble.
Time evolution of a generic statistical ensemble is described by the density matrix $\hat{\rho}(t)$ obeying the Liouville equation $i\hbar\partial_t\hat{\rho}(t)=[\hat{H}(t),\hat{\rho}(t)]$. Its formal solution can be written using the time-evolution operator $\hat{U}(t,t_0)=e^{-i\hat{H}(t-t_0)/\hbar}$ as $\hat{\rho}(t)=\hat{U}(t,t_0)\hat{\rho}(t_0)\hat{U}(t,t_0)^\dagger$. We can expand the density matrix at the initial time by the (many-body) eigenstates as $\hat{\rho}(t_0)=\sum_{\alpha\beta}\rho_{\alpha\beta}|u_\alpha\rangle\langle u_\beta|$, which leads to
\begin{align}
\hat{\rho}(t)=\sum_{\alpha\beta}\rho_{\alpha\beta}|u_\alpha\rangle\langle u_\beta| e^{-i(E_\alpha-E_\beta)(t-t_0)/\hbar}.
\end{align}
Here $E_\alpha$ and $|u_\alpha\rangle$ are eigenenergy and energy eigenstate of $\hat{H}$, respectively. Suppose that the energy eigenvalues have no degeneracy, long time average of the offdiagonal terms ($\alpha\neq\beta$) vanishes. This implies that, after a sufficiently long time, the density matrix $\hat{\rho}(t)$ is expected to be indistinguishable from the ensemble diagonal in the energy eigenstates (diagonal ensemble), \cite{Rigol2008}
\begin{align}
\hat{\rho}_\text{diag}=\sum_{\alpha}\rho_{\alpha\alpha}|u_\alpha\rangle\langle u_\alpha|.
\end{align}
Note that whether $\hat{\rho}_\text{diag}$ describes thermal equilibrium is not evident at this point.
Recall that we can expand the time-evolution operator of periodically-driven systems by Floquet states as Eq.~(\ref{eq:tevol-floquet}). If we focus on the stroboscopic time $t=t_0+nT$ with $n$ being an integer, the above argument can be straightforwardly applied by replacing $E_\alpha$ and $u_\alpha$ by the quasienergy $\epsilon_{\alpha}$ and periodic part of the Floquet state $|v_\alpha(t_0)\rangle$. Namely, the steady state of the periodically-driven system is expected to be characterized by an ensemble diagonal in the Floquet states, with coefficients $\rho_{\alpha\alpha}$ depending on the initial condition.

The diagonal ensemble indeed describes thermal equilibrium when the (static) Hamiltonian $\hat{H}$ satisfies a condition called the eigenstate thermalization hypothesis (ETH),\cite{Deutsch1991,Srednicki1994,Rigol2008} which is believed to hold for generic nonintegrable systems. Each energy eigenstate of the system with ETH is indistinguishable from thermal equilibrium, as long as local observables in the thermodynamic limit matter. Then the diagonal ensemble describes thermal equilibrium regardless of the initial condition $\rho_{\alpha\alpha}$, as long as the energy fluctuation is small enough in the initial state.
Namely, for periodically-driven systems, whether the thermal equilibrium of the effective Hamiltonian $\hat{F}$ is realized should depend on whether it satisfies ETH.

However, generic periodically driven nonintegrable many-body systems turn out to be quite different from static many-body systems, and show the ETH in a very unusual fashion.
Floquet states of such nonintegrable driven system have featureless expectation values corresponding to the infinite temperature for any local observable, which is called Floquet ETH.\cite{DAlessio2014,Lazarides2014}
Indeed, from the viewpoint of thermodynamics, the infinite temperature state is a natural consequence when an isolated system is continuously driven externally.

\begin{figure}
    \centering
    \includegraphics[width=\linewidth]{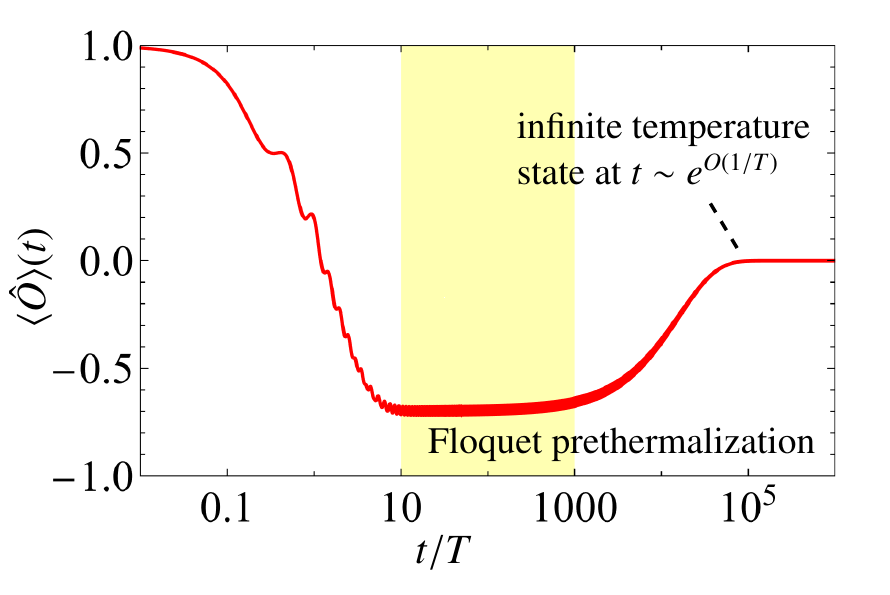}
    \caption{(Color Online)
    Schematic picture of the long-time behavior of an isolated periodically-driven system with high-frequency driving. The system first evolves to a steady state synchronized with external driving. This steady state is characterized by the Floquet effective Hamiltonian obtained by truncating the high-frequency expansion at an appropriate order. This relaxation to the steady state (realized in the highlighted region) is called Floquet prethermalization. The Floquet-prethermalized state sometimes exhibits nontrivial behavior distinct from the initial equilibrium state, which is schematically shown as a sign change in the expectation value of an operator $\hat{O}$ here. Generically, periodically-driven systems without coupling to an external environment heat up to an infinite temperature after a long time. In high-frequency driving, this is described as a deviation of the truncated expansion from the true Floquet Hamiltonian with featureless eigenstates. The heating phenomenon occurs in a 
    very slow time scale $\sim e^{O(1/T)}$ with $T=2\pi/\Omega$ being the period of driving.
    }
    \label{fig:thermal}
\end{figure}

While this might seem to contradict the fact that the high-frequency expansion yields a usual (many-body) effective Hamiltonian, it can be consistent due to the separation of time scale. Short-time dynamics can be approximated by the effective Hamiltonian obtained by the high-frequency expansion, while the inevitable truncation error leads to a slow heating toward infinite temperature. Many-body perturbative series are generically divergent, but have an optimal order where the truncation error becomes exponentially small.
For systems with an extensive upper bound for interaction energies, the time scale of the heating is shown to be bounded by $e^{O(\Omega)}$, i.e., the heating to the infinite temperature is quite slow.\cite{Kuwahara2016,Mori2016,Abanin2017} When there is such a separation of time scale, the system is expected to show first the thermalization toward the thermal equilibrium of the truncated effective Hamiltonian before heating up, which is called Floquet prethermalization. We show a schematic picture of this temporal structure in Fig.~\ref{fig:thermal}.

As we see above, we have to look at transient dynamics for exploring nontrivial states in many-body systems. When we have a small parameter other than $1/\Omega$ with a small truncation error, we can expect a similar prethermailzation behavior even when the high-frequency expansion is broken down. We show examples of Mott insulators with a large charge gap in Sec.~\ref{subsec:floquet-magnetic}.
On the other hand, when the many-body interaction is negligible, relaxation to the diagonal ensemble (without ETH) is expected. We can calculate such an ensemble, e.g., by explicitly setting $\hat{\rho}(t_0)$ as an equilibrium state without external drive, which corresponds to the dynamics where we suddenly apply the external field at $t=t_0$.\cite{Dehghani2016} 
Another way to avoid infinite heating and have nontrivial states is to consider an open dissipative setup, which is formulated mainly with nonequilibrium Green's function approach\cite{Aoki2014} and quantum master equation approach.\cite{Mori2023}

\section{Floquet Engineering of Geometrical Responses}

In this section, we review how geometrical responses introduced in Sec. 2 can be engineered in periodically driven systems, using the Floquet formalism described in Sec. 3.

\subsection{Graphene irradiated by circularly polarized light}\label{subsec:floquet-graphene}
Graphene illuminated by a circularly polarized light is the first example for 
Floquet engineering of the band topology.\cite{Oka2009,Kitagawa2011,Lindner2011,Oka2019,Rudner2020}
The tight-binding Hamiltonian for the irradiated graphene is given as 
\begin{equation}
\hat{H}(t)=\sum_{ij\sigma}t_{ij}e^{iq\bm{A}(t)\cdot\bm{R}_{ij}/\hbar}\hat{c}_{i\sigma}^{\dagger}\hat{c}_{j\sigma}\label{eq:peierls},
\end{equation}
where $t_{ij}$ is the hopping amplitude from site $j$ to $i$ in the absence of external fields, and
$\hat{c}_{i\sigma}$ is the annihilation operator for an electron with spin $\sigma\in\{\uparrow,\downarrow\}$ on the site at a position $\bm{R}_i$. We here only consider the nearest-neighbor hopping ($|\bm{R}_{ij}|=|\bm{R}_i-\bm{R}_j|=a$ with $a$ being the lattice constant) for simplicity.
The light electric field is introduced by the Peierls phase factor $e^{iq\bm{A}(t)\cdot\bm{R}_{ij}/\hbar}$, where $\bm{A}(t)$ is the vector potential. The electric field is represented as 
$\bm{E}(t)=-\partial_{t}\bm{A}(t)$. 
We adopt $\bm{A}(t)=(E_0/\Omega)(\cos\Omega t,\sin\Omega t,0)$ [i.e., $\bm{E}(t)=E_0(-\sin\Omega t,\cos\Omega t,0)$] for the vector potential of a monochromatic circularly polarized light, where $\Omega$ is the driving frequency and $E_0$ the amplitude of the electric field. 
We also introduce a dimensionless amplitude of the vector potential $A_0=qE_0a/\hbar\Omega$. Note that, even if $A_0$ is small, $E_0$ is not necessarily small when $\Omega$ is large.

Let us analyze this driven system using the scheme of the high-frequency expansion,\cite{Kitagawa2011,Jotzu2014,Eckardt2015,Mikami2016} assuming that the driving frequency is larger than the band width while it is much smaller than the energy gap from the bands not included in this model. The Peierls phase factor can be expanded into a Fourier series as
$t_{ij}e^{iq\bm{A}(t)\cdot\bm{R}_{ij}/\hbar}=\sum_{m=-\infty}^\infty t_{ij,m} e^{-im\Omega t}$ with
\begin{align}
t_{ij,m} &= t_{ij} i^m J_m(A_{ij})e^{im\phi_{ij}},\label{eq:peierls-fourier}
\end{align}
where $J_n$ is the $m$th Bessel function of the first kind,  $\bm{R}_{ij}=|\bm{R}_{ij}|(\cos\phi_{ij},\sin\phi_{ij},0)$, and $A_{ij}=A_0|\bm{R}_{ij}|/a$.
The leading order term of the effective Hamiltonian $\hat{H}_\text{eff}^{(0)}=\hat{H}_0$ [See Eq.~(\ref{eq:hfe-1st})] is then obtained as the Hamiltonian without the external field multiplied by the renormalization factor $J_0(A_0)$.
With this effect the band width of the irradiated system is reduced from that of equilibrium, and even vanishes when the amplitude $A_0$ is tuned to the zeros of the Bessel function.\cite{Dunlap1986} This phenomenon is called dynamical localization, and the Mott transition triggered by the dynamical localization is proposed and experimentally realized in (bosonic) ultra-cold atom systems.~\cite{Eckardt2005,Lignier2007,Eckardt2009}

\begin{figure}
    \centering
    \includegraphics[width=\linewidth]{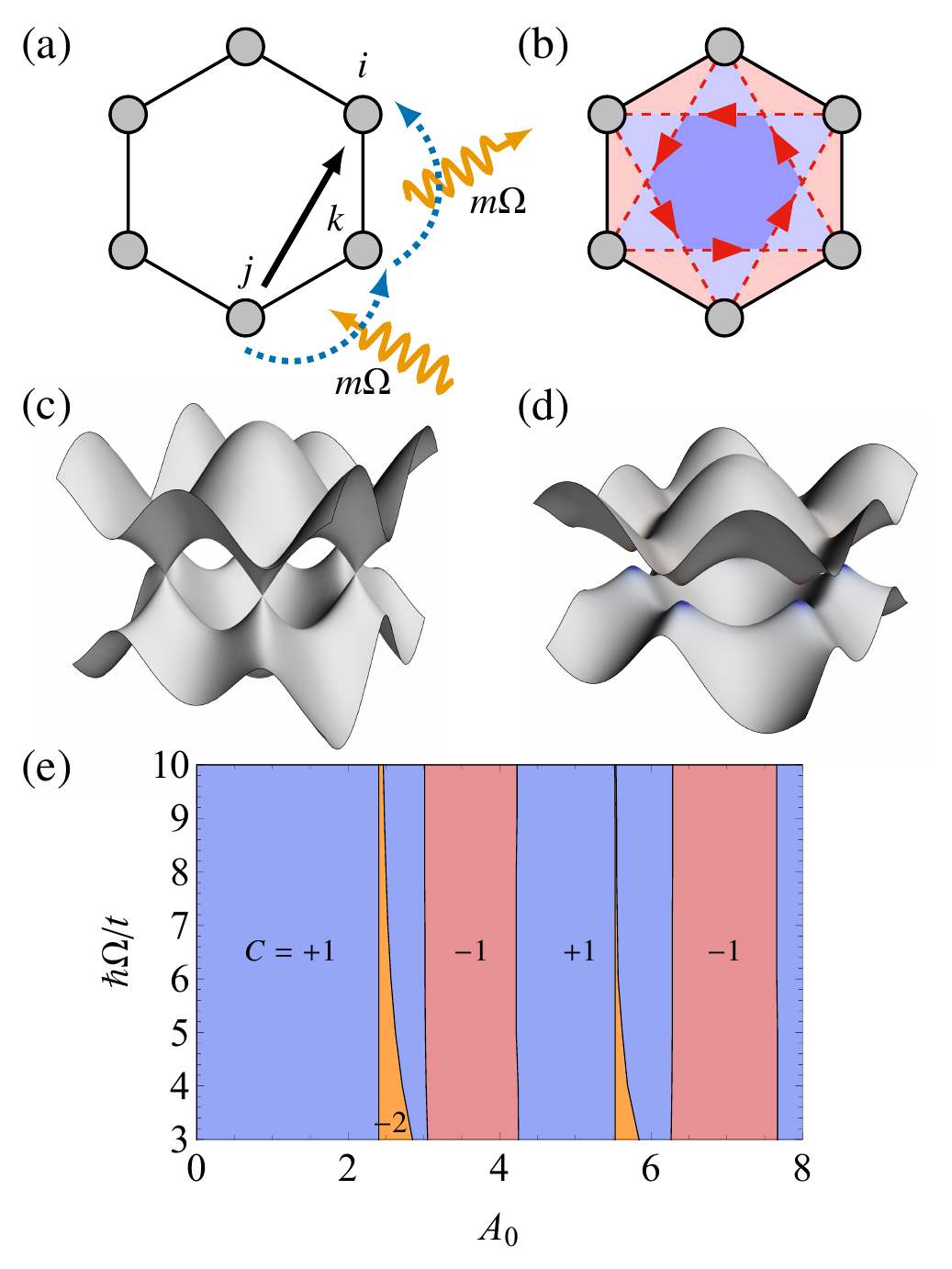}
    \caption{(Color Online)
    Graphene irradiated by a circularly-polarized light. 
    (a) Schematic picture of photo-induced next-nearest-neighbor hopping consisting of two steps with photon absorption and emission. (b) Haldane model with complex next-nearest-neighbor hopping, realized when the applied electric field is circularly polarized light.
    (c) band dispersion of the honeycomb lattice model in the absence of the external electric field.
    (d) band dispersion of the honeycomb lattice model when illuminated by a circularly-polarized light. Color represents the photo-induced Berry curvature, which takes the same sign for both K and K$'$ valleys.
    (e) Topological phase diagram as a function of the amplitude of the vector potential $A_0$ and driving frequency $\Omega$.\cite{Mikami2016} The distinct phases are characterized by the Chern number $C$. Topological phases with $C=+1,-1$, and $-2$ are colored blue, red, and orange, respectively.}
    \label{fig:graphene}
\end{figure}

The topological phase transition is triggered by the correction term $\hat{H}^{(1)}_\text{eff}$ [Eq.~(\ref{eq:hfe-2nd})]. In the present case, it yields the next-nearest-neighbor hopping $\kappa_{ij}$ represented as 
\begin{equation}
\kappa_{ij}=\sum_{k}\sum_{m\neq0} \frac{t_{ik,-m}t_{kj,m}}{m\hbar\Omega},\label{eq:photo-induced-hopping}
\end{equation}
which remarkably becomes imaginary when the field is chosen to be circularly polarized; Substituting the expression for the Fourier components (\ref{eq:peierls-fourier}) yields 
\begin{align}
\kappa_{ij}&=-i\sum_{k}\sum_{m=1}^\infty \frac{
2t_{ik}t_{kj}}{m\hbar\Omega} J_m(A_{ik})J_m(A_{jk})
 \sin m(\phi_{ik}-\phi_{jk})\label{eq:haldane-mass}\\
 &\simeq i\sum_{k}\frac{t_{ik}t_{kj}}{2\hbar\Omega}\left(\frac{qE_0}{\hbar\Omega}\right)^2 (\bm{R}_{ik}\times\bm{R}_{jk})_z,
\end{align}
which is nonvanishing in the honeycomb lattice geometry. We depict this perturbative process in Fig.~\ref{fig:graphene}(a).
The honeycomb lattice model with a complex next-nearest-neighbor hopping is known as the Haldane model~\cite{Haldane1988} with nontrivial Chern number [Schematically drawn in Fig.~\ref{fig:graphene}(b)],
which is originally recognized as a toy model but turned out to be feasible as a Floquet effective Hamiltonian as shown here.
The photo-induced next-nearest-neighbor hopping acts as a mass term for Dirac nodes of the original energy dispersion shown in Fig.~\ref{fig:graphene}(c), which leads to energy gaps with the nonzero Berry curvature with the same sign for K and K$'$ valleys as shown in Fig.~\ref{fig:graphene}(d), i.e., the system undergoes a phase transition to the Chern insulator.
This Floquet realization of the Haldane model is experimentally demonstrated in an ultra-cold atom system in Ref.~\citen{Jotzu2014}, and also explored in the Hall transport of graphene in Ref.~\citen{McIver2019}.

An advantage of the Floquet formalism is that there is no limitation on the strength of the field amplitude, as long as the photon energy $\hbar\Omega$ is taken large enough. For example, the photo-induced next-nearest-neighbor hopping Eq.~(\ref{eq:haldane-mass}) is an oscillating function of $A_0$, which is difficult to capture in a perturbative expansion with respect to $E_0$ (or $A_0$). Due to the oscillation, the Chern number of the system also exhibits multiple sign changes in the strong amplitude regime,\cite{Jotzu2014,Mikami2016} as shown in Fig.~\ref{fig:graphene}(e).

While the number of the edge states coincides with the Chern number of Floquet bands as in static problems, the distribution function for the edge states is not necessarily similar to equilibrium. Whether we can achieve a quantized Hall response in this setup is thus a nontrivial question. 
Specifically, we remark that for the present case with a $2\times2$ Hamiltonian in $k$ space, a no-go theorem for the topological phase transition is known for an isolated infinitely-large system, which originates from the smooth time evolution of the density matrix.\cite{DAlessio2015,Ge2017} However this constraint can be circumvented in various ways, e.g. by introducing edges, or taking account of nonunitary effects in an open-dissipative setup. 
In the pioneering work~\cite{Oka2009}, the Hall conductance is computed for an open-dissipative setup, but due to the low driving frequency no signature of quantization is observed. The possibility of the quantized transport is first discussed in Ref.~\citen{Kitagawa2011}, with an off-resonant driving. The (near) quantization is indeed numerically demonstrated in various setups~\cite{Dehghani2015,Mikami2016}.
In Ref.~\citen{Dehghani2015}, the coupling to phonons is considered with a rate equation approach, while in Ref.~\citen{Mikami2016}, influences of the electron-electron interaction and the coupling to a substrate are considered with the Floquet dynamical mean-field approach.
The quantized transport is also demonstrated in an isolated setup when the driving frequency is large enough~\cite{Dehghani2015,DAlessio2015}.

If we decrease the driving frequency $\Omega$ down to $\hbar\Omega<W$ with $W$ being the band width,
the quasienergy band undergoes a band folding in the energy direction, which cannot be captured by the high-frequency expansion. Then the folded band hybridizes with the other band, which results in a gap opening at the Floquet Brillouin zone boundary $\epsilon=\hbar\Omega/2$. This anticrossing turns out to accompany a change in the Chern number. \cite{Kundu2014,Piskunow2015,Mikami2016}
Of particular importance, after this topological transition, the Chern number can only distinguish the difference of the number of edge states for the energy gap at $\epsilon=0$ and $\epsilon=\hbar\Omega/2$. In an extreme case, these two gaps host chiral edge states even when the Chern numbers of the two bands are zero, which falls outside of the conventional topological classification for static systems due to the additional periodicity along the energy direction.
This topological phase with no static counterpart is called the anomalous Floquet topological insulator.\cite{Jiang2011,Rudner2013}
The classification of such Floquet anomalous topological phases are done in Ref.~\citen{Roy2017}.
This concept of novel quantum phases unique to periodically-driven systems is first proposed for gapless phases\cite{Kitagawa2010,Higashikawa2019}, and 
extended to (temporal) symmetry-protected topological insulators\cite{Else2016,Keyserlingk2016,Potter2016}, higher-order topological insulators\cite{Bomantara2019,Peng2019,Rodriguez2019,Huang2020}, and even the time crystals with spontaneous breaking of the discrete time-translational symmetry.\cite{Else2020}

\subsection{Application to magnetic systems}\label{subsec:floquet-magnetic}

Periodic driving is also used to engineer magnetic properties of materials.
A direct way to access magnetism by laser light is to use the magnetic field of the laser field. The coupling between spin degrees of freedom and the magnetic field is described by the Zeeman term $H_{\text{Z}}=-g\mu_B \bm{B}(t)\cdot\hat{\bm{S}}$.
When we apply a circularly polarized light, the static Hamiltonian in the Sambe space becomes block-diagonal due to the conservation of the angular momentum for spins and photons. One can obtain the effective static Hamiltonian in an exact manner, and the periodic field acts as a static magnetic field along the axial direction.\cite{Takayoshi2014} The strength of the effective field $\hbar\Omega$ is so large that one can draw the magnetization curve by continuously changing the driving frequency within the THz regime.\cite{Takayoshi2014-2}
Note that the vector potential adopted in the previous subsections is spatially uniform, and the magnetic field component is dropped (both for spin and orbital). 
The amplitude of the magnetic field $E_0/c$ is thus small and can be seen as a relativistic correction. 

On the other hand, one can also modulate spin-spin coupling solely by the electric field component.\cite{Mentink2014,Bukov2016,Kitamura2017,Claassen2017} This is because magnetism emerges in electronic systems due to the electron correlation, and the modulated correlated dynamics of electrons may affect spin structures.
To see this, let us consider the Hubbard model coupled to the laser electric field here,
\begin{align}
\hat{H}(t)&=\sum_{ij\sigma}t_{ij}e^{iq\bm{A}(t)\cdot\bm{R}_{ij}/\hbar}\hat{c}_{i\sigma}^{\dagger}\hat{c}_{j\sigma}+U\sum_{i}\hat{n}_{i\uparrow}\hat{n}_{i\downarrow}\label{eq:hubbard}\\
&=\hat{T}(t)+U\hat{D},
\end{align}
where $\hat{n}_{i\sigma}=\hat{c}_{i\sigma}^{\dagger}\hat{c}_{i\sigma}$. 
The low-energy physics of the half-filled Hubbard model is governed by the localized spin on each site when the repulsive onsite interaction $U$ is large enough and the system is in the Mott insulating phase.
In the absence of the external field, the Hubbard model is known to be mapped to the Heisenberg model
\begin{equation}
\hat{H}=\dfrac{1}{2}\sum_{ij}J_{ij}\left(\hat{\bm{S}}_{i}\cdot\hat{\bm{S}}_{j}-\frac{1}{4}\right)\label{eq:heisenberg-equil}
\end{equation}
with $J_{ij}=4|t_{ij}|^{2}/U$ and $\hat{\bm{S}}_{i}=1/2\sum_{\sigma\sigma^{\prime}}\hat{c}_{i\sigma}^{\dagger}\bm{\sigma}_{\sigma\sigma^{\prime}}\hat{c}_{i\sigma^{\prime}}$, by applying the degenerate perturbation theory. 
One can include the effect of the laser light field systematically 
by extending the perturbative method, with the help of the Floquet theory.

Let us briefly sketch the derivation of the effective low-energy Hamiltonian for the driven Hubbard model, using the method of the canonical transformation~\cite{Kitamura2017}. We here introduce a time-periodic unitary transformation of the Hamiltonian,
\begin{equation}
\hat{H}_{\text{SCE}}(t)=e^{i\hat{S}(t)}\hat{H}(t)e^{-i\hat{S}(t)}-e^{i\hat{S}(t)}i\hbar\frac{\partial}{\partial t}e^{-i\hat{S}(t)},\label{eq:canonical}
\end{equation}
with $\hat{S}(t)=\hat{S}(t+T)$.
One can easily check that $e^{i\hat{S}(t)}$ is represented as a static unitary matrix in the Sambe space representation, and does not change the quasienergy spectrum.
We choose $\hat{S}$ such that 
it cancels interaction terms that change the number of doubly-occupied sites  $\hat{D}=\sum_{i}\hat{n}_{i\uparrow}\hat{n}_{i\downarrow}$ (often called the doublon number). Namely, 
the transformed Hamiltonian $\hat{H}_{\text{SCE}}$ should satisfy $[\hat{H}_{\text{SCE}},\hat{D}]=0$, with which we can separate the low-energy excitations (i.e., those in the $\hat{D}=0$ subspace) from high-energy ones of $O(U)$.

\begin{figure}
    \centering
    \includegraphics[width=\linewidth]{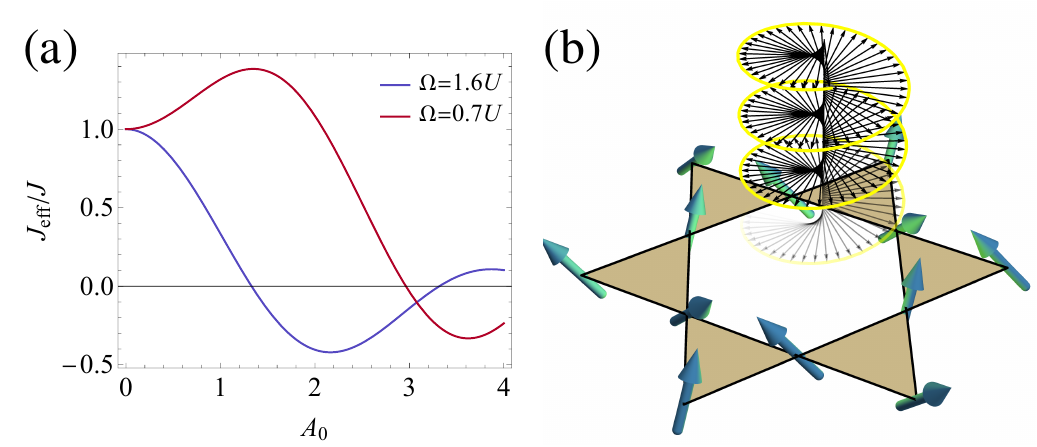}
    \caption{(Color Online)
    Modulation of spin correlation with the use of intense laser light.
    (a) Modulation of the kinetic exchange interaction, as a function of the amplitude of the vector potential $A_0$.\cite{Mentink2014}
    (b) An example of a magnetic order with nonzero scalar spin chirality on the kagome lattice. The scalar spin chirality can be induced when a circularly-polarized light is applied.\cite{Kitamura2017} }
    \label{fig:magnetic}
\end{figure}

Such a solution for $\hat{S}$ can be obtained order by order with respect to the hopping amplitude. The first-order term in Eq.~(\ref{eq:canonical}) reads
\begin{equation}
\hat{H}_{\text{SCE}}^{(1)}(t)=\hat{T}(t)+[i\hat{S}^{(1)}(t),U\hat{D}]-\hbar\frac{\partial\hat{S}^{(1)}}{\partial t},\label{eq:canonical-1st}
\end{equation}
where superscript indicates the order of the hopping amplitude. We set $\hat{S}^{(0)}=0$ here, because there is no term to cancel in the absence of the hopping term.
To determine $\hat{S}^{(1)}$ such that $[\hat{H}_{\text{SCE}}^{(1)},\hat{D}]=0$, 
it is convenient to decompose the hopping operator as $\hat{T}(t)=\hat{T}_{+1}(t)+\hat{T}_{0}(t)+\hat{T}_{-1}(t)$ with
\begin{equation}
\hat{T}_{+1}(t)
=\sum_{ij\sigma}t_{ij}e^{iq\bm{A}(t)\cdot\bm{R}_{ij}/\hbar}\hat{n}_{i\bar{\sigma}}\hat{c}_{i\sigma}^{\dagger}\hat{c}_{j\sigma}(1-\hat{n}_{j\bar{\sigma}})=\hat{T}_{-1}^{\dagger}(t),
\end{equation}
where $\bar{\uparrow}=\downarrow$, $\bar{\downarrow}=\uparrow$. Here the subscript represents the change in the doublon number, and the decomposed operators satisfy $[\hat{T}_d,\hat{D}]=-d\hat{T}_d$. By decomposing $\hat{S}^{(1)}=\sum_d\hat{S}^{(1)}_d$ in the same manner, we can compute the commutator in Eq.~(\ref{eq:canonical-1st}),
with which we obtain
\begin{equation}
\hat{S}^{(1)}(t)=-i\sum_{m=-\infty}^{\infty}\left(\dfrac{\hat{T}_{+1,m}}{U-m\hbar\Omega}-\dfrac{\hat{T}_{-1,m}}{U+m\hbar\Omega}\right)e^{-im\Omega t},
\end{equation}
and $\hat{H}_{\text{SCE}}^{(1)}(t)=\hat{T}_{0}(t)$,
where $\hat{T}_{d,m}$ is the Fourier component of the decomposed hopping operators $\hat{T}_d(t)=\sum_{m=-\infty}^{\infty}\hat{T}_{d,m}e^{-im\Omega t}$.
Proceeding to the higher orders in the same manner, we obtain
\begin{align}
\hat{H}_{\text{SCE}}(t) & =U\hat{D}+\hat{T}_{0}(t)+\sum_{nm}\left(\frac{[\hat{T}_{+1,m},\hat{T}_{-1,n-m}]}{2(U-m\hbar\Omega)}e^{-in\Omega t}+\text{H.c.}\right)
\end{align}
up to the second-order perturbation.
The block Hamiltonian for the $\hat{D}=0$ subspace describes the low-energy excitation of the driven Hubbard model. The equilibrium counterpart of the third term $[\hat{T}_{+1},\hat{T}_{-1}]/U$ corresponds to the 
exchange coupling Eq.~(\ref{eq:heisenberg-equil}), which implies that in the driven case we obtain a modulated spin coupling
\begin{align}
J_{ij}\to J_{ij}(t)= \sum_{nm}\frac{2\text{Re}[t_{ij,m}t_{ji,n-m}e^{-in\Omega t}]}{U-m\hbar\Omega}+(i\leftrightarrow j).\label{eq:heisenberg-modulated}
\end{align}
In particular, when we consider a monochromatic light with Eq.~(\ref{eq:peierls-fourier}) and perform the high-frequency expansion of this driven spin Hamiltonian, the spin-spin coupling is obtained as~\cite{Mentink2014}
\begin{align}
J_\text{eff} \simeq\overline{J_{ij}(t)}=\sum_{m}  \frac{4|t_{ij}|^2
J_{m}(A_{ij})^2
}{U-m\hbar\Omega}\simeq
\frac{4|t_{ij}|^2}{U}
+\frac{2|t_{ij}|^2
(qE_0R_{ij})^2 
}{U(U^2-\hbar^2\Omega^2)},\label{eq:heisenberg-modulated-weak}
\end{align}
which is shown in Fig.~\ref{fig:magnetic}(a).
Namely, in the presence of the strong electric field, the field-induced dynamics modulates the exchange coupling, which even undergoes a sign change when $U$ is slightly detuned from $\hbar\Omega$ and the contribution of $m=1$ term with an oscillating factor $J_1(A_0)$ becomes dominant.
This modulation of the spin coupling is experimentally explored in the ultracold atom systems, and in particular the development of a ferromagnetic correlation is demonstrated.~\cite{Gorg2018}

The interplay of the spin coupling modulation and the geometrical effect yields richer responses of quantum magnets.
Let us discuss how the electric polarization of the (driven) Hubbard model is related to its spin correlations. To this end, we first consider how the electric polarization operator $\hat{\bm{P}}=q\sum_{i\sigma}\bm{R}_i \hat{c}_{i\sigma}^\dagger\hat{c}_{i\sigma}$ can be represented in the low-energy theory. The effective polarization operator is obtained by the same unitary transformation as \cite{Morimoto-magnon21}
\begin{equation}
\hat{\bm{P}}_\text{spin}= \hat{P}_0e^{i\hat{S}(t)}\hat{\bm{P}}e^{-i\hat{S}(t)}\hat{P}_0,
\end{equation}
where $\hat{P}_0=\prod_i (1-\hat{n}_{i\uparrow}\hat{n}_{i\downarrow})$ is the projection operator to the spin space.
The leading-order term of the strong-coupling expansion is obtained at the second order as
\begin{align}
\hat{\bm{P}}_\text{spin}^{(2)}&=-\frac{1}{2}\sum_{ijnm}\frac{4\text{Re}[t_{ij,n}t_{ji,m}e^{-i(n+m)\Omega t}]}{(U+n\hbar\Omega)(U-m\hbar\Omega)}q\bm{R}_{ij}\left(\hat{\bm{S}}_{i}\cdot\hat{\bm{S}}_{j}-\frac{1}{4}\right)\\
&\simeq\frac{1}{2}\sum_{ij}\frac{8|t_{ij}|^{2}q\bm{E}(t)\cdot\bm{R}_{ij}}{ U(U^{2}-\hbar^2\Omega^{2})}q\bm{R}_{ij}\left(\hat{\bm{S}}_{i}\cdot\hat{\bm{S}}_{j}-\frac{1}{4}\right),
\end{align}
which is induced due to the external electric field.
A remarkable point here is that the polarization depends on the spin correlation $\bm{S}_{i}\cdot\bm{S}_{j}$. This originates from the fact that electrons can hop to neighboring sites only when they have an antiparallel spin configuration, which is also applied to the field-induced motion of electrons. The field-induced modulation of the Heisenberg coupling in Eq.~(\ref{eq:heisenberg-modulated-weak}) can be interpreted as a coupling between the field-induced polarization and the electric field, which has a nonzero time average.

It is interesting to see how the counterpart of an inherent polarization of Bloch electrons with a geometric origin appears in the low-energy structure of the Mott-insulating phase. 
Indeed we can find third-order terms that is nonvanishing in the zero-field limit, 
\begin{align}
\hat{\bm{P}}_{\text{spin}}^{(3)} & \simeq\sum_{ij}\frac{4|t_{ij}|^{2}V_{ij}}{U^{3}}q\bm{R}_{ij}\left(\hat{\bm{S}}_{i}\cdot\hat{\bm{S}}_{j}-\frac{1}{4}\right)\nonumber\\
 & -\sum_{ijk}\frac{4t_{ij}t_{jk}t_{ki}}{U^{3}}q\bm{R}_{ij}(\hat{\bm{S}}_{i}-\hat{\bm{S}}_{j})\cdot\hat{\bm{S}}_{k},
\end{align}
where $V_{ij}=t_{ii}-t_{jj}$ is the difference of the onsite potential. Here we have assumed that bare hopping $t_{ij}$ is real. Correspondingly, the spin Hamiltonian at the third-order in the hopping amplitude for a weak field case reads
\begin{align}
H_{\text{SCE}}^{(3)}&\simeq-\frac{\bm{E}(t)\cdot\hat{\bm{P}}_{\text{spin}}^{(3)}}{(1-\hbar^2\Omega^{2}/U^{2})^{2}}.
\end{align}
The electric polarization proportional to $\bm{S}_{i}\cdot\bm{S}_{j}$ and its magnetoelectric coupling is known to show up in multiferroic materials\cite{Tokura2014} and called the exchange striction mechanism,\cite{Jia2006,Katsura2009} which is intensively studied along with another mechanism with nonzero vector chirality $\bm{S}_{i}\times\bm{S}_{j}$ known as the inverse Dzyaloshinskii-Moriya effect. \cite{Katsura2005,Mostovoy2006}
Floquet analysis here provides a microscopic derivation of the magnetoelectric coupling between the polarization (magnetostriction) intrinsic to the Hubbard-type electronic model and the external electric field, both in dc and ac cases beyond the linear response regime.
Floquet engineering of spin systems via this direct coupling between spin and ac electric field is discussed in Ref.~\citen{Sato2016} with a phenomenological coupling constant.

As well as the electric polarization, the electric current can also be expressed in terms of the spin degree of freedom.\cite{Bulaevskii2008}
The local current operator on the bond $ij$, defined as $\hat{\bm{J}}_{ij}=-i(q/\hbar)t_{ij}\bm{R}_{ij}\sum_{\sigma}c_{i\sigma}^{\dagger}c_{j\sigma}e^{iq\bm{A}(t)\cdot\bm{R}_{ij}/\hbar}+\text{H.c.}$, is written in the low-energy description as $\hat{\bm{J}}_{\text{spin},ij}= \hat{P}_0e^{i\hat{S}(t)}\hat{\bm{J}}_{ij}e^{-i\hat{S}(t)}\hat{P}_0$. Its zero field expression in the third order of the strong coupling expansion reads
\begin{equation}
\hat{\bm{J}}_{\text{spin},ij}^{(3)}=\sum_{k}\frac{6qt_{ij}t_{jk}t_{ki}}{\hbar U^{2}}\bm{R}_{ij}(\hat{\bm{S}}_{i}\times\hat{\bm{S}}_{j})\cdot\hat{\bm{S}}_{k}+\text{H.c.}
\end{equation}
Here, $(\hat{\bm{S}}_{i}\times\hat{\bm{S}}_{j})\cdot\hat{\bm{S}}_{k}$ is so-called scalar spin chirality [depicted in Fig.~\ref{fig:magnetic}(b)]. While the net current (summed over $ij$) vanishes identically, nonzero loop current and resultant orbital magnetic moment should be induced in the presence of the scalar spin chirality.
Indeed, when the hopping amplitude $t_{ij}$ becomes complex in the presence of a (static) magnetic flux represented by a spatially-nonuniform Peierls phase,\cite{footnotesk2}
the effective spin Hamiltonian acquires a coupling between magnetic flux and orbital moment (scalar chirality) as~\cite{Sen1995,Motrunich2006}
\begin{equation}
\hat{H}_\text{spin}^{(3)}  =\dfrac{1}{6}\sum_{ijk}J_{\chi}(\hat{\bm{S}}_{i}\times\hat{\bm{S}}_{j})\cdot\hat{\bm{S}}_{k},\quad
J_{\chi}  =\frac{24\text{Im}[t_{ij}t_{jk}t_{ki}]}{U^{2}}.
\end{equation}

In the noninteracting Bloch electron systems, 
the Hall response arises even without the (net) magnetic field, 
when the system has nonzero Berry curvature due to a nontrivial geometry,
which, as we have seen, can be controlled using circularly-polarized laser light.
This implies that we can control the scalar spin chirality using circularly-polarized light as well. Indeed, if we proceed to the fourth order in the hopping amplitude, we obtain~\cite{Kitamura2017,Kitamura2022}
\begin{align}
J_{\chi}&\simeq-\sum_{h}\frac{2t_{ij}t_{jk}t_{kh}t_{hi}\hbar\Omega(7U^{2}-3\hbar^2\Omega^{2})}{U^{2}(U^{2}-\hbar^2\Omega^{2})^{3}}(qE_0)^2
(\bm{R}_{ih}\times\bm{R}_{hk})_z\label{eq:chirality}
\end{align}
in the leading order of the field amplitude $E_0$. 

The scalar spin chirality has a strong influence on the geometric property of materials in various ways. Since it manifests a solid angle subtended by three spins, it tends to induce noncoplanar spin structures typified by the magnetic skyrmion.
Electrons propagating on the background of spin texture with nonzero scalar chirality acquire the Berry phase, yielding the topological Hall effect.\cite{Nagaosa2013}
Noncoplanar spin structures often accompany a nontrivial geometric structure on the magnon excitations, which leads to a Hall transport of magnons.
In Ref.~\citen{Owerre2017}, the emergent scalar spin chirality term in periodically-driven systems is discussed in terms of the Aharonov-Casher phase of magnons, and it is shown that the scalar chirality term makes the magnon band topological.

The scalar spin chirality yields nontrivial topological properties even in the absence of the magnetic order. The chiral spin liquid phase with fractional excitations is proposed to emerge in the kagome lattice model in the presence of the scalar spin chirality term,\cite{Bauer2014} and its dynamical control in terms of Floquet engineering is also discussed~\cite{Claassen2017}.

\subsection{Application to superconductors}

Dynamical control of superconductors experimentally undergoes intense study.
It includes excitation of Higgs modes by an intense THz laser \cite{Matsunaga2014,Shimano2020} and an ultrafast phase transition from normal to superconducting state \cite{Cavalleri2017}, which triggered various theoretical studies of superconductivity based on the Floquet formalism.

\begin{figure}
    \centering
    \includegraphics[width=\linewidth]{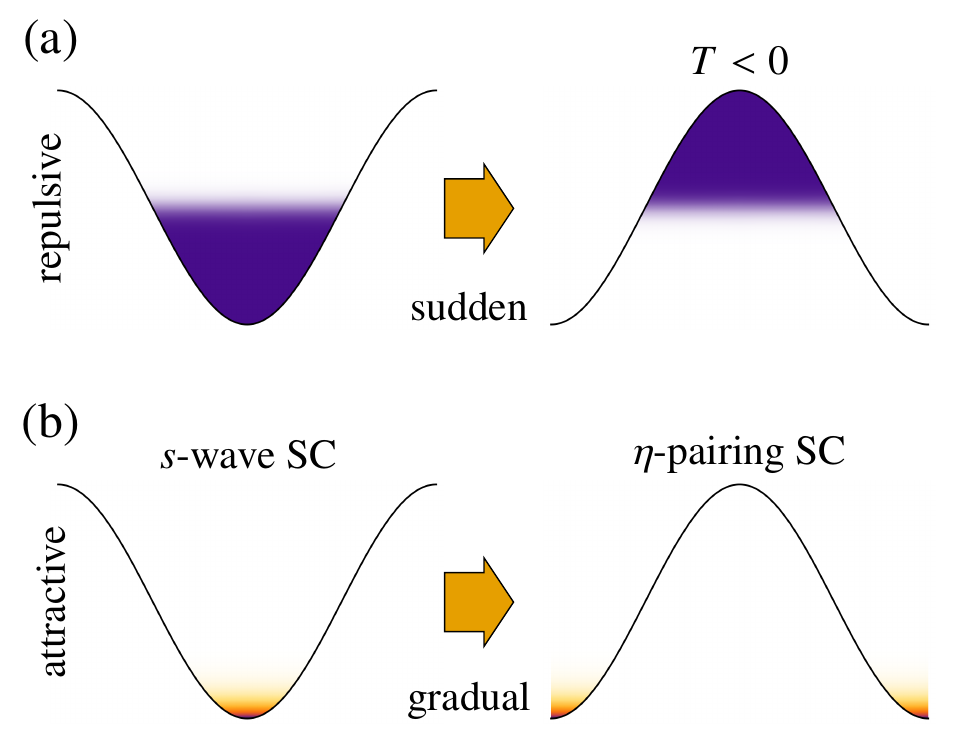}
    \caption{(Color Online)
    (a) Schematic picture of the band flipping for a weakly-repulsive Hubbard model, where the kinetic term is dominant and band picture is suitable.
    When intense laser light is applied and the renormalization factor $J_0(A_0)$ has a sign change, the energy dispersion is inverted. If the field is applied suddenly, an inverted population with $T<0$ should be realized, which can be reinterpreted as $T>0$ with a weakly-attractive interaction.~\cite{Tsuji2011}
    (b) Schematic picture of the band flipping for a strongly-attractive Hubbard model, where the electrons form Cooper pairs in real space and behave as bosons with a charge $-2e$. The ground state exhibits $s$-wave superconductivity. After the sign change of the renormalization factor $J_0(2A_0)$ induced by an intense light, the ground state changes to the $\eta$-pairing superconductivity, where the center-of-mass momentum of the Cooper pairs locates at the corner of the Brillouin zone. This state can be realized when the field amplitude is gradually increased.~\cite{Kitamura2016}
    }
    \label{fig:flip}
\end{figure}

Here let us feature two examples of superconducting transition based on the ``dynamical band flipping" mechanism,\cite{Tsuji2011,Kitamura2016} which can be understood in terms of the engineering of the Floquet effective Hamiltonian.
We first consider the driven Hubbard model Eq.~(\ref{eq:hubbard}) with a weak repulsive interaction. If we consider the high-frequency expansion (assuming $U\lesssim t\ll\hbar\Omega$), the leading-order effect is the renormalization of the hopping term by a factor $J_0(A_0)$, as we have seen in Sec.~\ref{subsec:floquet-graphene}. 
If we apply intense laser light stronger than that required for the dynamical localization ($2.4<A_0<5.5$), the renormalization factor becomes negative and the band dispersion is flipped. As the hopping term dominates the interaction term in the present case, 
when we apply the external field in an abrupt manner, we obtain an inverted population with a negative temperature.~\cite{Tsuji2011} This situation is equivalent to the positive temperature state with an attractive interaction, so that we can expect to have a superconducting transition if heating and dissipation effects are negligible [See Fig.~\ref{fig:flip}(a)].

Another example focuses on the (cubic lattice) Hubbard model with a strong attractive interaction,~\cite{Kitamura2016} where the electrons form a bosonic molecule in real space. 
As such molecular bosons have a charge $-2e$, they are expected to undergo the dynamical localization and dynamical band flipping described by the renormalization factor $J_0(2A_0)$ in the strong field regime. In the band-flipped regime, the ground state of the effective Hamiltonian is the Bose-Einstein condensate with the center-of-mass momentum at the corner of the Brillouin zone, which is called the $\eta$-pairing state, which is schematically depicted in Fig.~\ref{fig:flip}(b).
This phenomenon can be quantitatively described by the strong coupling expansion introduced in Sec.~\ref{subsec:floquet-magnetic}. In the attractive case, we can derive the low-energy Hamiltonian by focusing on the subspace with maximum doublon number $\hat{D}$,
which is spanned by the pseudospin composed of doubly-occupied and empty sites,
\begin{gather}
\hat{\eta}_{i}^{x}+i\hat{\eta}_{i}^{y}=(-1)^{i}\hat{c}_{i\uparrow}\hat{c}_{i\downarrow},\quad\hat{\eta}_{i}^{z}=\dfrac{1-\hat{n}_{i}}{2}.
\end{gather}
with $\hat{n}_{i}=\hat{n}_{i\uparrow}+\hat{n}_{i\downarrow}$
The second-order effective Hamiltonian is obtained as the XXZ model, with the $J_\text{eff}^z$ given by Eq.~(\ref{eq:heisenberg-modulated-weak}) and $J_{\text{eff}}^x$ given by
\begin{align}
J_{\text{eff}}^x&= \sum_{m=-\infty}^{\infty}(-1)^m\frac{4\text{Re}t_{ij}^2J_m(A_0)^2}{U-m\hbar\Omega}
\simeq\frac{4\text{Re}t_{ij}^2}{U}J_0(2A_0),
\end{align}
where the last expression is obtained by neglecting $\hbar\Omega/U$.
Behind this modulation of the $\eta$-spin interaction, the hidden symmetry of the Hubbard model plays a crucial role. The mapping called Shiba transformation
$\hat{c}_{i\uparrow}\to\hat{c}_{i\uparrow}^{\dagger}(-1)^{i}$
defined for bipartite lattices [$(-1)^{i}$ takes $+1$ ($-1$) for $A$ ($B$) sublatice] transforms the hopping and interaction terms as
\begin{gather}
\sum_{ij}t_{ij}\hat{c}_{i\uparrow}^{\dagger}\hat{c}_{j\uparrow}\to\sum_{ij}t_{ji}\hat{c}_{i\uparrow}^{\dagger}\hat{c}_{j\uparrow},\\
U\sum_{i}\left(\hat{n}_{i\uparrow}-\frac{1}{2}\right)\left(\hat{n}_{i\downarrow}-\frac{1}{2}\right)\to-U\sum_{i}\left(\hat{n}_{i\uparrow}-\frac{1}{2}\right)\left(\hat{n}_{i\downarrow}-\frac{1}{2}\right),
\end{gather}
which makes the undriven (half-filled) repulsive and attractive Hubbard model equivalent.
Under this mapping the spin operator is transformed as $\hat{\bm{S}}_i\to\hat{\bm{\eta}}_i$, which clarifies the hidden SU(2) symmetry for the $\eta$-spin. 
A remarkable point here is that the presence of the Peierls phase factor in the driven case leads to the breaking of the above equivalence under $t_{ij}=t_{ji}$,
and thus the system is no longer $\eta$-SU(2) symmetric.
This feature plays a key role in realizing the $\eta$-pairing state in an isolated system, where the conservation of the quasienergy obstructs the relaxation from the original $s$-wave superconducting state when we simply invert the band structure for molecular bosons. When we gradually increase the field amplitude $A_0$, the system first enters into the region where the charge-ordered state is the ground state, which arises due to the broken $\eta$-SU(2). Development of the charge order during the time evolution leads to the emergence of the $\eta$-pairing in the final state.~\cite{Kitamura2016}
Thanks to the coupling between electric fields and $\eta$ spins discussed above,
even in repulsive cases one can observe a development of an $\eta$-pairing correlation in the time evolution and the steady state of the driven Hubbard model. \cite{Kaneko2019,Li2020,Peronaci2020}

\begin{figure*}[t]
    \centering
    \includegraphics[width=\linewidth]{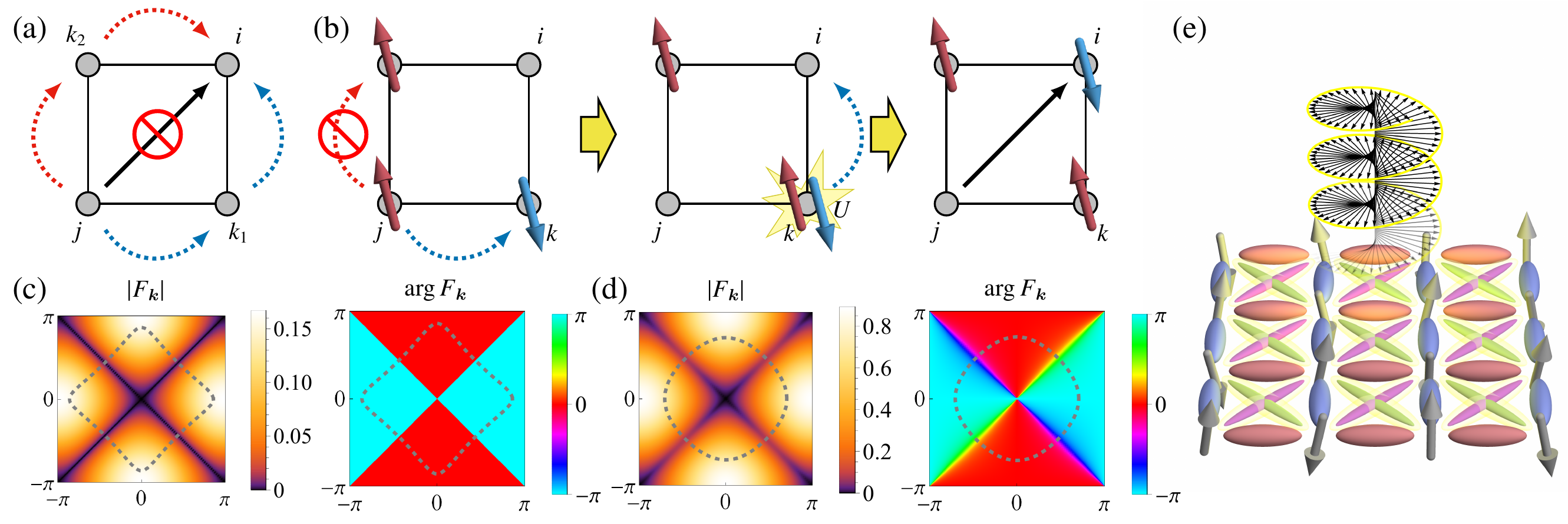}
    \caption{(Color Online)
    Cuprate superconductor irradiated by a circularly-polarized light. 
    (a) Cancellation of the photo-induced imaginary next-nearest-neighbor hopping in the square lattice.
    (b) A perturbation process in the strong coupling expansion for the three-site term.
    (c) Gap function $F_{\bm{k}}$ for the $t$--$J$ model in the absence of the external electric field. Dashed curve represents the Fermi surface at the critical temperature. The gap function has nodal lines along $k_x=\pm k_y$.
    (d) Gap function $F_{\bm{k}}$ for the $t$--$J$ model illuminated by a circularly-polarized light. The energy spectrum becomes fully-gapped with a nontrivial winding of the phase factor of the gap function. (e) Schematic picture of the photo-induced chiral $d+id$-wave superconductivity.\cite{Kitamura2022} 
    Arrows represent localized spins with an antiferromagnetic fluctuation on the square lattice. Red and blue bonds represent $d_{x^2-y^2}$-wave pairing amplitude with positive and negative signs.
    Illumination of circularly-polarized light induces imaginary diagonal bonds (magenta and green) with $d_{xy}$-wave symmetry.   
    Adapted from Ref.~\citen{Kitamura2022}. \copyright[2022]{CC BY 4.0.}
    }
    \label{fig:tsc}
\end{figure*}

The concept of Floquet engineering is also applied to the engineering of the band topology to realize topological superconductivity.
In early works~\cite{Ezawa2015,Takasan2017}, topological superconductivity realized by the nontrivial geometry of the normal band structure is proposed.
Reference \citen{Ezawa2015} discusses the proximity effect of $s$-wave superconductor on the Haldane model discussed in Sec.~\ref{subsec:floquet-graphene}, while Ref.~\citen{Takasan2017} considers the doped repulsive Hubbard model with $d$-wave superconductivity illuminated by the circularly-polarized light. In the latter case with a usual square lattice geometry (with application to cuprates in mind), the correction term of the high-frequency expansion $\hat{H}_\text{eff}^{(1)}=[\hat{H}_{-m},\hat{H}_m]/2m\hbar\Omega$ identically vanishes:
While the correction term can be rewritten as the next-nearest-neighbor hopping with the same form as Eq.~(\ref{eq:haldane-mass}), summation on the intermediate site $k$ results in a cancellation of all the contributions in the square lattice, as depicted in Fig.~\ref{fig:tsc}(a).
The correction term becomes finite when one considers a Rashba spin-orbit coupling $\lambda$ on the thin-film geometry, which leads to the ($k$-dependent) Zeeman coupling inducing the topological gap with an order of $\sim A_0^2\lambda^2/\hbar\Omega$.

In the above examples, the topological phase transition is triggered by the change in the normal band structure, while the symmetry of the superconducting gap function remains the same as the equilibrium state. While the topological phase transition were realized more easily if we could modify the symmetry of the gap function directly,
it is challenging because the charge-neutral gap function has no direct coupling to the laser electric field due to the inherent particle-hole symmetry.\cite{Pekker2015}

The technique of the strong coupling expansion introduced in Sec.~\ref{subsec:floquet-magnetic} is also useful here for investigating the modulation in the pairing interaction of the correlated superconductivity,~\cite{Kitamura2022} as we can systematically describe the influence of the external field on the low-energy structure even in the nonlinear regime.
With the use of the strong-coupling expansion, in the undriven case, it is known that the Hubbard model can be mapped to the $t$--$J$ model~\cite{Ogata2008}
\begin{align}
\hat{H}_{t\text{--}J}&=
\sum_{ij\sigma}t_{ij}\hat{P}_{0}\hat{c}_{i\sigma}^{\dagger}\hat{c}_{j\sigma}\hat{P}_{0}+\frac{1}{2}\sum_{ij}J_{ij}\left(\hat{\bm{S}}_{i}\cdot\hat{\bm{S}}_{j}-\frac{\hat{n}_i\hat{n}_j}{4}\right)\nonumber\\
& +\sum_{ijk\sigma\sigma^{\prime}}\Gamma_{ij;k}\hat{P}_{0}\hat{c}_{i\sigma}^{\dagger}\hat{c}_{j\sigma^{\prime}}\hat{P}_{0}\left(\bm{\sigma}_{\sigma\sigma^{\prime}}\cdot\hat{\bm{S}}_{k}-\frac{\delta_{\sigma\sigma^{\prime}}}{2}\hat{n}_{k}\right),
\end{align}
where $J_{ij}=4|t_{ij}|^2/U$ is the same as the half-filled case, while $\Gamma_{ij;k}=t_{ik}t_{kj}/U$ is the so-called three-site term, which describes a two-step hopping process of a (singly-occupied) electron from $j$ to $i$ site (a hole from $i$ to $j$) passing through $k$ site, which is shown in Fig.~\ref{fig:tsc}(b). 
Due to the strong correlation and the Pauli exclusion, such a hopping process is only allowed when the electrons on the path have an antiparallel spin configuration. 

When we apply a circularly-polarized light on the doped Hubbard model, we obtain a modified $t$--$J$ model as a Floquet effective Hamiltonian, in a similar manner to the half-filled case discussed in Sec.~\ref{subsec:floquet-magnetic}; The hopping amplitude $t_{ij}$ acquires a renormalization factor $J_0(A_{ij})$ leading to the dynamical localization, while the exchange interaction $J_{ij}$ is modulated as Eq.~(\ref{eq:heisenberg-modulated-weak}).
The three-site term $\Gamma_{ij;k}$ absent at half filling is also modulated, whose expression reads
\begin{align}
\Gamma_{ij;k} & =\frac{t_{ik,0}t_{kj,0}}{U}+\sum_{m\neq0}\dfrac{t_{ik,-m}t_{kj,m}}{m\hbar\Omega(1-m\hbar\Omega/U)}\label{eq:correlated-hopping}\\
 & =t_{ik}t_{kj}\left(\dfrac{J_{0}(A_{ik})J_{0}(A_{kj})}{U}+\sum_{m\neq0}\dfrac{J_{m}(A_{ik})J_{m}(A_{kj})e^{im(\phi_{jk}-\phi_{ik})}}{m\hbar\Omega(1-m\hbar\Omega/U)}\right).
\end{align}
Recall that this expression is quite similar to the photo-induced hopping Eq.~(\ref{eq:photo-induced-hopping}) in noninteracting systems, which becomes complex when the external field is circularly polarized. 
Indeed, the amplitude of the three-site term becomes complex in the present case as
\begin{align}
\text{Im}\Gamma_{ij;k} & \simeq \dfrac{t_{ik}t_{kj}}{2\hbar\Omega(1-\hbar^2\Omega^2/U^2)}\left(\frac{qE_0}{\hbar\Omega}\right)^2(\bm{R}_{ik}\times \bm{R}_{jk})_z.
\end{align}
A remarkable point here is that this term even survives in the square lattice case, which contrasts with the photo-induced hopping in noninteracting systems;
The noninteracting square lattice case with no configuration dependence undergoes 
the cancellation of contribution upon the summation over site $k$ [See Fig.~\ref{fig:tsc}(a)]. 
In this sense, the present time-reversal symmetry-breaking term is peculiar to correlated systems.

Also, at the fourth-order perturbation we obtain the scalar spin chirality term with broken time-reversal symmetry, as in the half-filled case. While the leading-order expression Eq.~(\ref{eq:chirality}) vanishes for the square lattice case after the summation over site $h$, we have nonzero chiral coupling $\sim E_0^4 t^2t^{\prime2}/\hbar\Omega U^6$ when considering next-nearest-neighbor hopping $t^\prime$ and consider higher orders in $E_0$.

These time-reversal broken interactions modify
the pairing interaction $V_{\bm{k},\bm{k}^\prime}$ for the gap equation of the superconductivity  $F_{\bm{k}}=\sum_{\bm{k}^\prime}V_{\bm{k},\bm{k}^\prime}\langle\hat{c}_{-\bm{k}^\prime\downarrow}\hat{c}_{\bm{k}^\prime\uparrow}\rangle$.
In particular, for the square lattice Hubbard model describing high-$T_c$ cuprates, we obtain
\begin{align}
V_{\bm{k},\bm{k}^\prime} &\simeq
-\frac{3}{N}[J_\text{eff}f_{\bm{k}}f_{\bm{k}^\prime}
+4i(2\delta \gamma+J_\chi) (g_{\bm{k}}f_{\bm{k}^\prime}
-f_{\bm{k}}g_{\bm{k}^\prime})]
\end{align}
within the Gutzwiller approximation, 
where $f_{\bm{k}}=\cos k_x-\cos k_y$, $g_{\bm{k}}=2\sin k_x\sin k_y$ are the form factor, $\delta$ is the doping level and $\gamma = \text{Im}\,(\Gamma_{i-x,i;\,i+y}-\Gamma_{i-x-y,i+x;\,i})$ is the difference of two three-site terms $\propto t t^\prime$. Here $J_\chi$ denotes the coupling constant for $\sum_i(\hat{\bm{S}}_i\times\hat{\bm{S}}_{i+y})\cdot\hat{\bm{S}}_{i+x}$.
Due to the emergent chiral component, the pairing symmetry of the superconductivity changes from $d_{x^2-y^2}$-wave [shown in Fig.~\ref{fig:tsc}(c)] to $d_{x^2-y^2}+id_{xy}$-wave [Fig.~\ref{fig:tsc}(d)], which is schematically depicted in Fig.~\ref{fig:tsc}(e).
Due to this change in the gap function, the Bogoliubov-de Gennes Hamiltonian within the Gutzwiller approximation becomes fully gapped along with a nontrivial Chern number.

\section{Geometrical Nonlinear Optical Responses \label{sec: NLOR}}
In this section, we review nonlinear optical responses that originate from a nontrivial geometry of Bloch electrons in solids.
We explain two such examples from bulk photovoltaic effects: shift current and quantized circular photogalvanic effect in Weyl semimetals.

\subsection{Bulk photovoltaic effects in solids}

The bulk photovoltaic effect (BPVE) is a phenomenon where light irradiation to bulk crystals leads to a dc current generation. \cite{Boyd,Bloembergen,Sturman}
The BPVE is described by the second order nonlinear current response as
\begin{align}
    J_\mu=\sigma^{(2)}_{\mu\alpha\beta}(\omega) E_\alpha(\omega)E_\beta(-\omega),
    \label{eq: sigma (2)}
\end{align}
where $J_\mu$ is the dc current along the $\mu$ direction, $E_\alpha(\omega)$ is the Fourier component of the electric field along the $\alpha$ direction with the frequency $\omega$, and
$\sigma^{(2)}_{\mu\alpha\beta}(\omega)$ is the nonlinear conductivity tensor.
In inversion symmetric systems,
the nonlinear conductivity should vanish.
Specifically, $J_\mu$ is odd and $E_\alpha$ is even under the inversion symmetry, and hence the left hand side of \eqref{eq: sigma (2)} changes the sign under the inversion while the right hand side does not, which requires $\sigma^{(2)}_{\mu\alpha\beta}(\omega)=0$.
In a conventional photovoltaic effect in p-n junctions for photovoltaic effects, the inversion symmetry is broken due to the artificial structure at the interface of p- and n-type semiconductors. 
In contrast, BPVE does not require such an artificial heterostructure and may generally appear in bulk crystals with broken inversion symmetry.

There are several mechanisms for BPVEs.
One mechanism is a shift current which arises from a shift of electron wave packet upon optical transition and has a close relationship with ``the modern theory of electric polarization" as we will explain in detail. \cite{Baltz,Belinicher82,Sipe,Young-Rappe,Young-Zheng-Rappe,Cook17,Morimoto-Nagaosa16,Tan16,Sotome19,Burger19,Hatada20}
Another mechanism is a ballistic current where a ballistic motion of photoexcited electrons carries the electric current according to their group velocities. \cite{Belinicher-Sturman80,Belinicher-Sturman88}
This nonlinear current response is also called the injection current when it occurs from a purely electronic process. \cite{Sipe,deJuan17,Orenstein21} 
Ballistic current also takes place due to scattering processes for photoexcited carriers with impurities and phonons.
The term "ballistic current" is sometimes used in a narrow sense when the current flow by a ballistic motion of electrons arises from a scattering process of the photoexcited electron, distinguishing it from the injection current as a purely electronic process. 

When the system preserves the time reversal symmetry (TRS), application of linearly polarized light leads to shift current and ballistic current (involving scattering processes),
while the injection current is forbidden since the band structure becomes symmetric in the momentum space and contributions of the group velocity cancel for time reversal pairs of electrons.
Circularly polarized light can also generate injection current since circular polarization effectively breaks TRS and induces an imbalance of photocarriers in the momentum space.
The BPVE under a circularly polarized light is called the circular photogalvanic effect (CPGE). \cite{orenstein2021topology}

\subsection{Shift current as a geometric bulk photovoltaic effect}

\begin{figure}
    \centering
    \includegraphics[width=\linewidth]{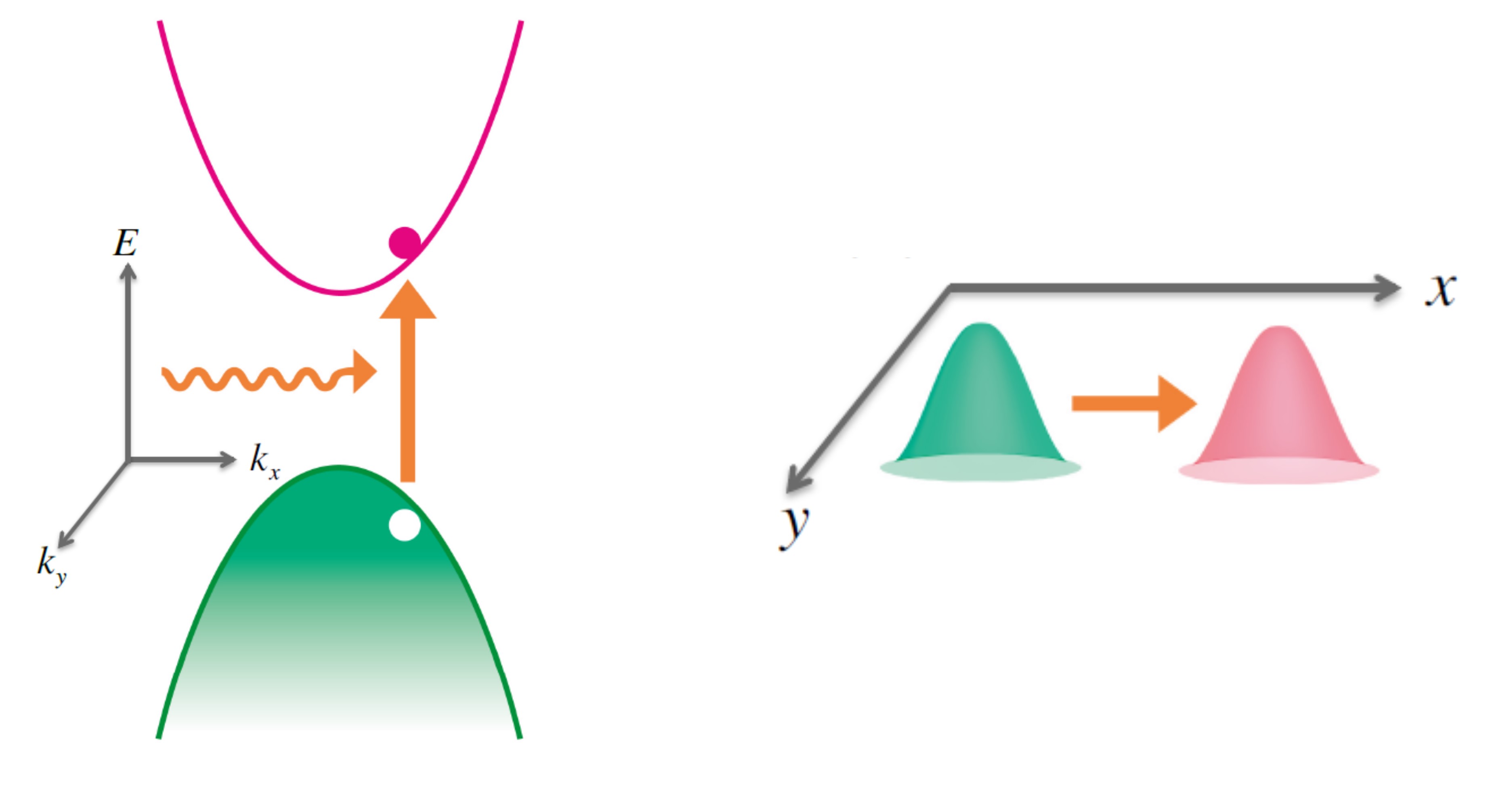}
    \caption{(Color Online)
    Schematic picture of shift current.
    Upon an interband optical transition,
    the wave packet shifts in real space according to the difference of the intracell coordinates of the valence and conduction bands, leading to a nonzero electric polarization of the photoexcited electron-hole pair.
    In the steady state under light irradiation, the continuous creation of photocarriers linearly increases the electric polarization in the system which results in dc current generation.
    Taken from Ref.~\citen{Kun17}. \copyright[2017]{American Physical Society.}
    }
    \label{fig: wavepacket shift}
\end{figure}

The shift current is one mechanism of the BPVE in solids that originates from the nonzero electric polarization of photoexcited electron-hole pairs.  
When the photon energy of the incident light is above the band gap, an interband transition between valence and conduction bands generally leads to the shift of the wave packet in the inversion broken materials and generates nonzero electric polarization, as illustrated in Fig.~\ref{fig: wavepacket shift}.
Under the continuous illumination of light, electron-hole pairs are created constantly in time and the electric polarization $P$ due to the shift of wave packets grows linearly in time.
A general relationship between the current and the polarization $J=dP/dt$ indicates that such a linear growth of $P$ results in dc current generation, which is the shift current.

\begin{figure}
    \centering
    \includegraphics[width=1.0\linewidth]{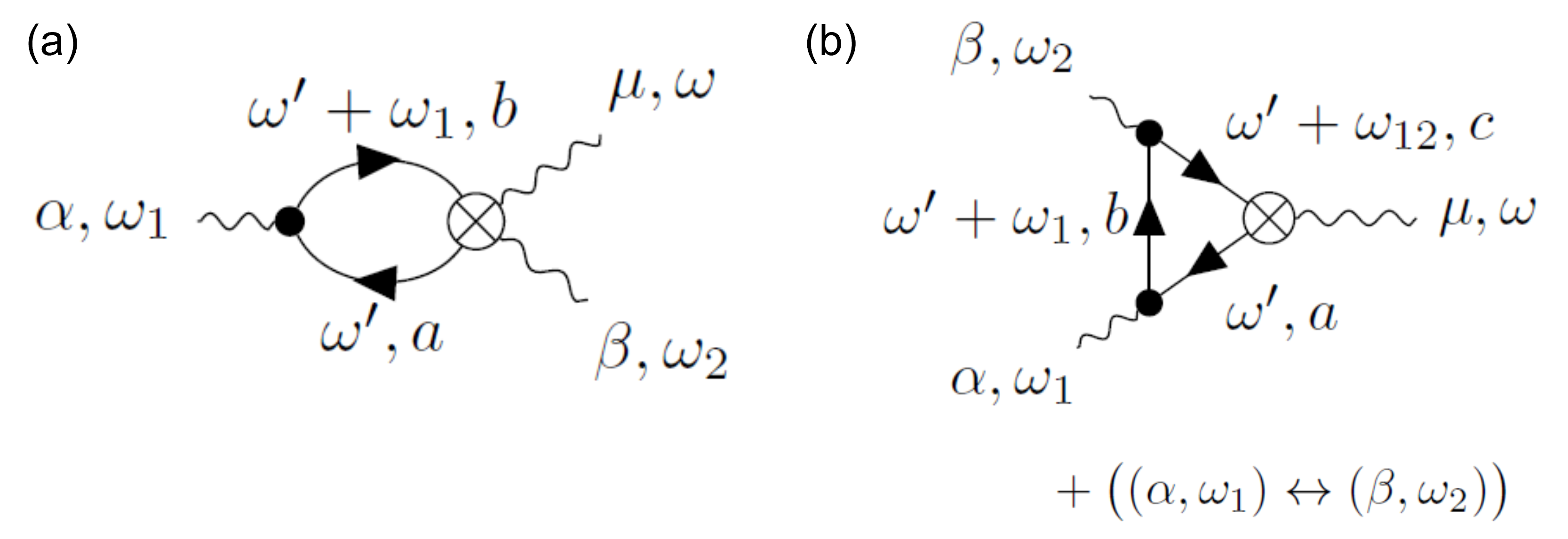}
    \caption{Diagrams contributing to the shift current. 
    (a) Bubble diagram of paramagnetic current $v$ and diamagnetic current $\partial_\mu v$
    (b) Triangle diagram of paramagnetic current $v$.
    Taken from Ref.~\citen{Parker19}. \copyright[2019]{American Physical Society.}
    }
    \label{fig: shift diagram}
\end{figure}

Microscopic derivations of the shift current are given by a second order perturbation theory~\cite{Baltz,Sipe,Young-Rappe,Parker19}. 
From the standard diagrammatic approach, the nonlinear conductivity for shift current is contributed by the bubble diagram and the triangle diagram (Fig.~\ref{fig: shift diagram}), which gives for the case of linearly polarized light with $\alpha=\beta$,
\cite{Parker19}
\begin{align}
    \sigma^{(2)}_{\mu\alpha\alpha}(\omega) &=
    \frac{q^3}{\hbar^3\omega^2} \int \frac{dk}{(2\pi)^d} \Bigg[ 
    \sum_{a,b}
    \frac{f_{ab}(\partial_{k_\mu}v_\alpha)_{ab}v_{\alpha,ba} }{\hbar \omega - \epsilon_{ba}+i\delta}
    \n
    &
    + \sum_{a,b,c\neq a}
    \frac{v_{\mu,ac}v_{\alpha,cb}v_{\alpha,ba}}{\epsilon_{ac}}
    \left( \frac{f_{ab}}{\hbar\omega-\epsilon_{ba}+i\delta}-\frac{f_{cb}}{\hbar\omega-\epsilon_{bc}-i\delta} \right) 
    \bigg] 
    \n
    &+ (\omega \leftrightarrow -\omega),
    \label{eq: sigma2 perturbation}
\end{align}
where $q$ is the electron charge, the subscripts $a,b,c$ label the energy bands, $f_{ab}=f_a-f_b$ with the Fermi distribution $f_a$ for the state $a$, $\epsilon_{ab}=\epsilon_{a}-\epsilon_{b}$ with the energy $\epsilon_{a}$ of the state $a$,
and $\delta$ is the energy broadening.
Here, we defined the velocity operator $v_\mu=\partial_{k_\mu} H(k)$ along the $\mu$ direction and its matrix element 
$v_{\mu,ab}=\bra{u_a} v_\mu \ket{u_b}$.
The first term in the integrand corresponds to the bubble diagram of the paramagnetic current $v_\alpha$ and the diamagnetic current $\partial_{k_\mu} v_\alpha$, and the second term corresponds to the triangle diagram of $v_\alpha$.
We note that the $k$ derivative in defining the diamagnetic current $\partial_{k_\mu} v_\alpha$ should be taken using the basis of wave functions that have no $k$ dependence, such as the site representation in tight binding models. 

For time reversal symmetric systems,
the resonant contribution to the photocurrent in Eq.~\eqref{eq: sigma2 perturbation} is given by the shift current contribution, which can be rewritten concisely as \cite{Sipe,Young-Rappe}
\begin{align}
J_{shift}^{\mu\alpha\alpha}
&= \frac{\pi q^3}{\hbar^3 \omega^2} |E_\alpha(\omega)|^2
\sum_{ab}\int[dk] f_{ab}|v_{\alpha,ab}|^2 R_{\mu\alpha,ba} \delta(\hbar\omega-\epsilon_{ba}),
\label{eq: shift current}
\end{align}
with the shift vector
\begin{align}
    R_{\mu\alpha,ba} &=
\left[
\frac{\partial}{\partial k_\mu}\mathrm{Im}(\log v_{\alpha,ab})
+ a_{\mu,b}- a_{\mu,a}
\right].
\label{eq:Rk}
\end{align}
The shift vector includes the difference of the Berry connections ($a_{\mu,a}=i\braket{u_a|\partial_{k_\mu} |u_a}$) for the two bands involved in the optical transition, accompanied by the $k$ derivative of the phase of the interband velocity matrix element that ensures the gauge invariance of $R$.
Since the Berry connection is the intracell coordinate of the Bloch electrons, the shift vector essentially measures the shift of the wave packet upon optical transition between the two bands.
The square of the interband velocity matrix element $|v_{ab}|^2$ describes the rate of the optical transition. 
The combination of these two quantities determines the increase of polarization $P$ in the steady state and the resulting dc current generation.
In the nonequilibrium steady state under the light irradiation, the polarization $P$ does not increase to infinity; the increase of the polarization is balanced with the current generation through the electrodes attached to the sample and the decrease of the polarization in the local recombination processes of electron-hole pairs.
Equation \eqref{eq: shift current} only describes the initial process ($dP/dt$) and how much out of $dP/dt$ can be extracted as a dc current out of the sample heavily depends on the extrinsic processes including a scattering effect of photocarriers within the sample and an interface effect between the sample and the electrode. \cite{nakamura2017shift,nakamura2018impact}

Because the shift current arises from the shift of an electron wave packet with a geometric origin, the amount of photocurrent does not depend on the relaxation rate of the carriers, which contrasts with the conventional photocurrent from the ballistic motion of the carriers. 
Indeed, it was reported that shift current in a ferroelectric SbSI shows significant robustness against the disorder level of the sample. \cite{Hatada20}
Also, the geometric nature of the shift current leads to the absence of shot noise for the photocurrent \cite{Morimoto-IV18},
which suggests that the shift current photovoltaics is potentially suitable for application to photodetectors.

Equation \eqref{eq: shift current} obtained from the diagram in Fig.~\ref{fig: shift diagram} is based on the perturbation with respect to the vector potential $A$ coupling to the current in the system, which is called the velocity gauge. \cite{Kraut79,Baltz}
Another approach is called the length gauge where one considers the perturbation with respect to the electric field $E$ coupling to the electric polarization or the position operator of electrons that is represented by the covariant derivative. \cite{Aversa95,Sipe}
These two approaches are known to be equivalent as far as contributions from all the energy bands are incorporated.
The proof of the equivalence involves the usage of sum rules on the matrix elements of velocity operators, which is nontrivial. \cite{Ventura17,Passos,Taghizadeh18}
The velocity gauge often shows an artificial divergence in the dc limit if all the contributions are not properly taken into account, whereas the length gauge has the advantage that it is easy to take the dc limit without such divergences.
Also, a truncation of bands in the velocity gauge tends to cause more numerical errors for BPVE in the optical region, due to neglected contributions from intermediate states (included in the second term on the right-hand side in Eq.~\eqref{eq: sigma2 perturbation}).\cite{taghizadeh17}
That said, the velocity gauge has the advantage that it has a clear diagrammatic interpretation and is easy to incorporate electron correlation effects based on the diagrammatic treatment \cite{Parker19,Morimoto-magnon19}.

\subsection{First principles calculations of shift current}
Shift current has been actively studied based on the first principles calculations. \cite{Nastos10,Young-Rappe,Young-Zheng-Rappe,Tan16,dai2022recent}
Young and Rappe \cite{Young-Rappe} performed a first principles calculation of shift current for BaTiO$_3$ from the expression involving the shift vector, 
and made a comparison with the earlier experimental result of BPVE in BaTiO$_3$, \cite{koch1975bulk}
succeeding to reproduce characteristic frequency and polarization dependencies of the observed photocurrent. 
Later the first principles calculation of BPVE in BaTiO$_3$ has been further elaborated by including an electron correlation effect by the GW approximation \cite{Fei20}
and incorporating the ballistic current contributions from the asymmetric scattering of photoexcited carriers arising from an electron-phonon interaction \cite{Dai21-prl} and an electron-hole interaction \cite{Dai21-prb}.

Bismuth ferrite (BiFeO$_3$) has a low band gap and is attracting interest as a potential photovoltaic material with a large open circuit voltage \cite{yang2010above}.
Shift current response of BiFeO$_3$ has been studied by the first principle calculation with the DFT+U method \cite{Young-Zheng-Rappe}.
The first principle calculation of BPVE in single-layer monochalcogenides GeS also predicted a large effective three-dimensional shift current ($\simeq$ 100  $\mu$A/V$^2$). \cite{Rangel17}

Application of a strain to materials can induce polarity to the system and may enhance shift current responses.
Such strain effect on the BPVE was named "piezophotovoltaic effect" and an ab initio study was performed for the transition metal dichalcogenide 2H-MoS$_2$. \cite{schankler2021large}
Experimentally, an introduction of 0.2\% tensile strain to 3R-MoS$_2$ was reported to induce a giant BPVE which is over two orders of magnitude larger than that in an unstrained sample. \cite{dong2022giant} 

\subsection{Floquet approach to nonlinear optical effects}
\begin{figure}
    \centering
    \includegraphics[width=\linewidth]{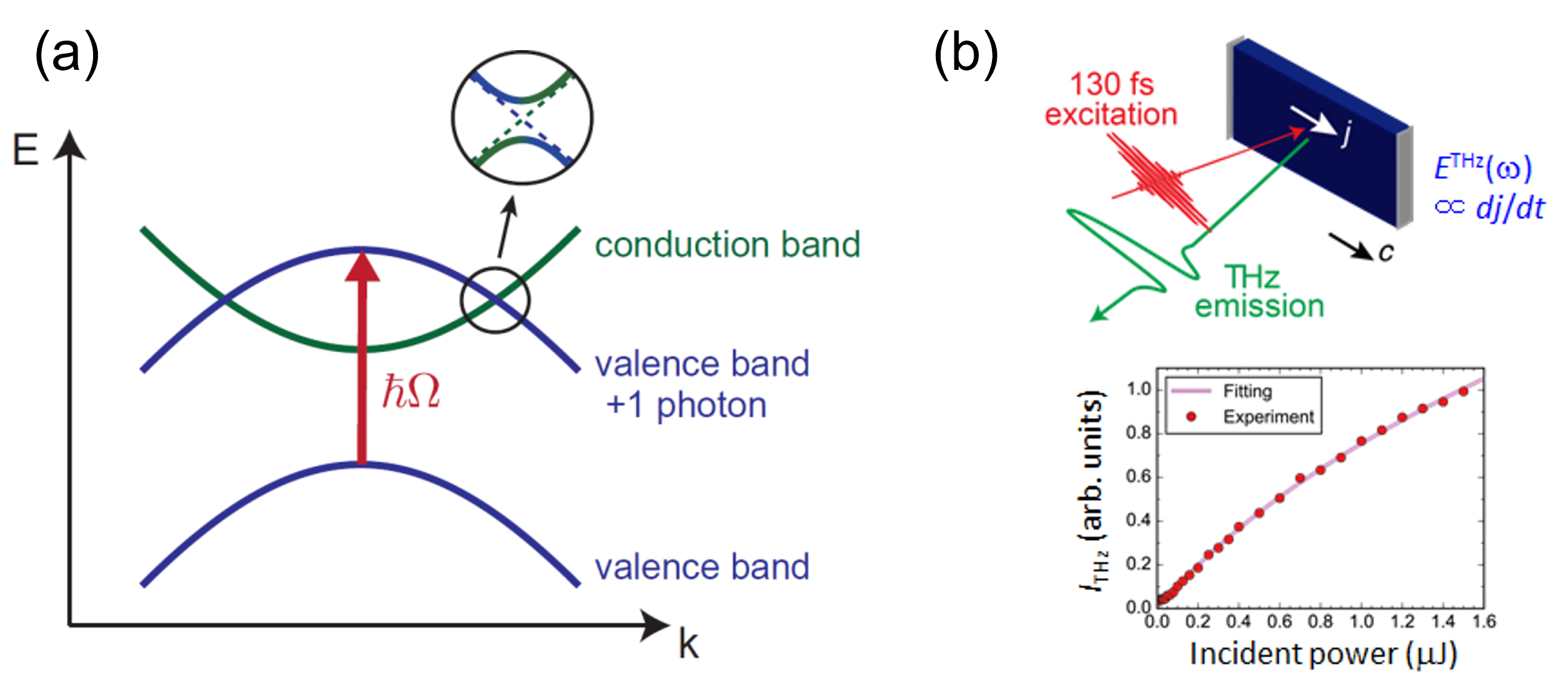}
    \caption{(Color Online)
    (a) Floquet formalism for nonlinear optical effects. 
    Taken from Ref.~\citen{Morimoto-Nagaosa16}. \copyright [2016] CC BY-NC 4.0. 
    (b) THz spectroscopy of shift current in ferroelectrics SbSI. The saturation effect of the photocurrent is well fitted by the Floquet formula for shift current.
    Taken from Ref.~\citen{Sotome19}. \copyright [2019] CC BY-NC-ND 4.0.
    }
    \label{fig:floquet}
\end{figure}

Floquet theory that treats periodically driven systems provides an alternative theoretical approach to the nonlinear optical effects,
since an electron system subjected to an external electric field of the monochromatic light can be regarded as a periodically driven system \cite{Morimoto-Nagaosa16}.
In the Floquet formalism for optical responses, there appear copies of energy bands that are shifted by the photon energy in the energy direction and are labeled by the Floquet indices.
When an optical transition takes place, a valence band (band 1) attached with one photon and a conduction band (band 2) exhibit an anticrossing in the Floquet formalism as shown in Fig.~\ref{fig:floquet}(a), which is described by the Floquet Hamiltonian (with a convention $e=\hbar=1$), 
\begin{align}
H_F&=
\begin{pmatrix}
\epsilon_1+ \Omega & -iA^* v_{12} \\
iA v_{21} & \epsilon_2
\end{pmatrix},
\label{eq: HF}
\end{align}
where $A$ and $\Omega$ are the vector potential and frequency of the monochromatic light.
In this formalism, the associated dc current operator $v_F$ is given by
\begin{align}
    v_F&=
    \begin{pmatrix}
        v_{11} & -iA^*\left( \frac{\partial v}{\partial k} \right)_{12} \\
        iA\left( \frac{\partial v}{\partial k} \right)_{21} & v_{22}
    \end{pmatrix},
\end{align}
where $v_{ij}$ is the matrix element of the velocity operator in the unperturbed system.

Some remarks are in order here. In Eq.~(\ref{eq: HF}),
the matrix representation is in the Bloch wave function basis. Therefore, the basis itself is dependent on $k$. This leads to the difference between 
$\biggl( \frac{\partial v}{\partial k} \biggr)_{12}$
and $\frac{\partial v_{12}}{\partial k}$ as given 
by 
\begin{align}
\biggl( \frac{\partial v}{\partial k} \biggr)_{12}   =  \frac{\partial v_{12}}{\partial k} - \braket{ \partial_k u_1 |v| u_2 } - \braket{ u_1 |v | \partial_k u_2 },
\end{align}
which is further transformed into 
\begin{align}
\biggl( \frac{\partial v}{\partial k} \biggr)_{12} 
&= \frac{\partial v_{12}}{\partial k} - 
i (a_1-a_2) v_{12}     \nonumber \\
&+ \braket{ \partial_k u_1 |u_2 } v_{22} - v_{11} \braket{ u_1 | \partial_k u_2 }    \nonumber \\
&= 
v_{12} \biggl[ \partial_k \log v_{12}
- i (a_1 - a_2) + 
\frac{v_{11} - v_{22}}{\epsilon_1 - \epsilon_2} \biggr].
\label{eq: dv_12}
\end{align}
In the above equation, the imaginary part of 
$[\cdots]$ in the last line corresponds to the shift vector in Eq.~\eqref{eq:Rk} as \cite{footnotetm1}
\begin{align}
    \mathrm{Im}\left[\frac{1}{v_{12}} \biggl( \frac{\partial v}{\partial k} \biggr)_{12} \right]
    &=\mathrm{Im}\left[\partial_k \log v_{12} +a_2 -a_1 \right]
    =R_{21},
\end{align}
quantifying the positional shift of the wave packet from band 1 to band 2.
Also, here it is assumed that $\ket{u_1}\bra{u_1} + \ket{u_2}\bra{u_2} =1$,
i.e., $\ket{u_1}$ and $\ket{u_2}$ form the basis of the Hilbert space of the present problem. 
This is an approximation assuming that this 
two-dimensional space describes the low energy 
physics relevant to the optical transition with the frequency $\Omega$.
Namely, the truncation has been done at the level of $k$-independent basis not after the diagonalization obtaining the Bloch wave functions.
Therefore, the Berry connections $a_1$ and $a_2$ are also approximate ones different from the exact ones obtained by the first-principles calculation using the full basis without the truncation.

Floquet theory combined with Keldysh Green's function technique concisely reproduces an expression for the shift current \eqref{eq: shift current} for two band systems \cite{Morimoto-Nagaosa16}.
This is essentially achieved by computing the expectation value of $v_F$ for the nonequilibrium steady states given by $H_F$.
The Floquet approach has the advantage that it can incorporate a nonperturbative effect in $E$.
Specifically, the result of the Floquet approach for the shift current includes an additional factor 
$(\Gamma/2)/\sqrt{E^2 |v_{12}|^2/\Omega^2 + \Gamma^2/4}$
with energy broadening $\Gamma$.
This additional factor describes the crossover of the shift current response from $J\propto E^2$ in the weak intensity region to $J\propto E\Gamma$ in the high intensity region.
Such crossover is a saturation effect of the current response for large light intensity $E^2$ which comes from the stimulated emission of the photoexcited electrons in the conduction band.
This saturation effect is a hallmark of the shift current mechanism of the BPVE and is observed in the THz spectroscopy of the shift current in ferroelectrics SbSI (Fig.~\ref{fig:floquet}(b)) \cite{Sotome19}.

\subsection{Shift current in systems with electron interactions}
Shift current in inversion broken semiconductors originates from nonzero electric polarization of photoexcited electron-hole pairs.
In this regard, when an elementary excitation that is optically allowed supports nonzero electric polarization  $P$, it is expected that the continuous creation of such elementary excitation with photoirradiation can also lead to photocurrent generation from a similar shift current mechanism.

\begin{figure}
    \centering
    \includegraphics[width=0.8\linewidth]{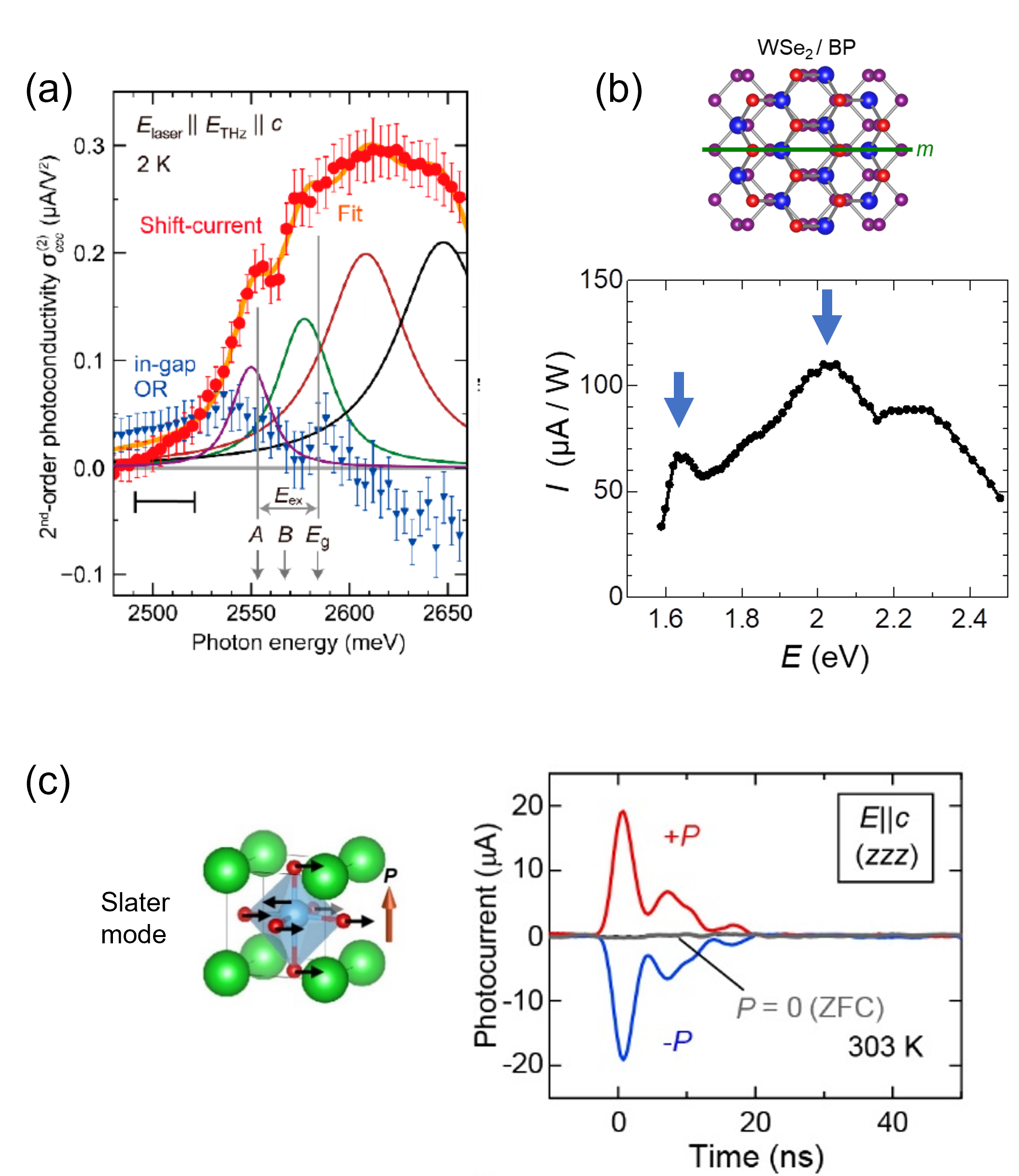}
    \caption{(Color Online)
    Shift current in systems with electron interactions.  (a) Photoconductivity of a semiconductor CdS. Exciton shift current appears as a peak structure at the A exciton resonance. Taken from Ref.~\citen{Sotome21}. \copyright[2021]{American Physical Society.} 
    (b) Photocurrent of a polar interface of WSe$_2$ and black phosphorus. The photocurrent shows an enhancement at the exciton resonances of WSe$_2$ (blue arrows) due to the exciton shift current contribution. Taken from Ref.~\citen{Akamatsu21}. \copyright[2021]{The American Association for the Advancement of Science.}
    (c) Photocurrent by soft phonon excitations in a ferroelectric BaTiO$_3$. 
    Irradiation of a THz light induces photocurrent that depends on the polarity due to the shift current mechanism.
    Ref.~\citen{Okamura22}. \copyright[2022]{CC BY-NC-ND 4.0.}
    }
    \label{fig:exciton shift current}
\end{figure}

An exciton is a bound state of photoexcited electron and hole. 
Since the wave packets for valence and conduction electrons have a positional shift in inversion broken crystals, an exciton in those crystals generally supports nonzero $P$.
Thus it was theoretically predicted that photoexcitation of excitons can support dc current generation due to shift current \cite{Morimoto-exciton16,Chan21}.
For shallow excitons, the shift current is expressed with the shift vector at the band gap as \cite{Morimoto-exciton16,Sotome21}
\begin{align}
    J(\omega)\simeq \frac{2e^3}{\hbar^2\omega^2} |E|^2 (\delta k)^3 |v_{12}(k=0)|^2 R(k=0) \frac{\Gamma}{(\omega-\omega_\mathrm{ex})^2+\Gamma^2},
\end{align}
for light frequency $\omega$,
where the band gap is assumed to be at $k=0$, 
$(\delta k)^3$ is the small volume in the Brillouin zone within the energy range of exciton binding energy above the band gap, and $\Gamma$ is the scattering rate of the electron-hole pairs.
The shift current of excitons is observed in a semiconductor CdS in the THz spectroscopy \cite{Sotome21}.
In Fig.~\ref{fig:exciton shift current}(a), the photocurrent spectrum shows a peak structure at the A exciton resonance below the band gap, which corresponds to the exciton shift current. 
A polar interface of transition metal dichalcogenide WSe$_2$ and black phosphorus was also shown to support exciton shift current \cite{Akamatsu21}.
In Fig.~\ref{fig:exciton shift current}(b), blue arrows indicate the exciton energy of WSe$_2$, where an enhancement of photocurrent at exciton resonances arises from the exciton shift current.


In magnets with inversion symmetry breaking, magnetic excitations accompany electric polarization due to the multiferroic nature, known as electromagnons. \cite{Pimenov06,Aguilar09,Takahashi12, kubacka2014large}
Since electromangons are optically accessible, 
nonzero electric polarization of electromagnons implies a possibility of photocurrent generation with optical excitation of electromagnons from the shift current mechanism.
Such magnon shift current was predicted by a toy model of multiferroic material based on a Hubbard model with spin orbit coupling \cite{Morimoto-magnon19}.
Magnon shift current has also been studied within the framework of a spin Hamiltonian and a magnon representation, where coupling to the external electric field is incorporated
by an electric polarization from an exchange striction mechanism or an inverse Dzyaloshinskii–Moriya interaction 
\cite{Morimoto-magnon21}.
Photoexcitation of magnons is also predicted to support dc spin current as a second order process \cite{Ishizuka2019,Ikeda2019}.
In the case of collinear antiferromagnets,
the spin current from the shift current mechanism can be expressed by a shift vector of magnons with Berry connection of magnon bands \cite{fujiwara2022nonlinear}.

Phonons can also carry electric polarization through an electron-phonon coupling when the crystal breaks inversion symmetry.
Thus a photocurrent by phonon excitation is also expected from the shift current mechanism, 
although the phonons are charge neutral and their energy scale is much smaller than the electronic band gap.
Such phonon shift current has been observed in a typical ferroelectric BaTiO$_3$ \cite{Okamura22}.
In the ferroelectric phase of BaTiO$_3$, the optical phonon mode perpendicular to the polarization direction (Slater mode) appears in the terahertz (THz) regime with an enhanced oscillator strength due to softening.
Photoexcitation of the soft phonon mode with a THz light induces dc current generation which changes its direction depending on the polarity of the sample [Fig.~\ref{fig:exciton shift current}(c)].
The observed photocurrent is consistent with the theoretical calculations based on the shift current mechanism.
The Glass coefficient of the observed phonon shift current is as large as $\simeq 1\times 10^{-8}$ cm/V, which is comparable with that for electronic excitations in BaTiO$_3$, 
while the energy scale of the soft phonon mode is three orders of magnitude smaller than the electronic band gap.

Elementary excitations such as excitons, magnons, and phonons are charge neutral and appear below the electronic band gap.
It is highly nontrivial how the charge-neutral excitations generated by photoirradiation turn into net charge current.
The polarization generated at the light spot may initially induce photocurrent by capacitive coupling to the electrodes. 
At a later time, an increase in electric polarization within the sample
can generate charge carriers with extrinsic processes such as a collision of excitations and scattering at impurity centers, which eventually carry the net charge current compensating for the increasing electric polarization in the sample.
Namely, the expression for the shift current only captures the initial process of an increase of $P$ by the photoexcitations,
while the real photocurrent is expected to be partially canceled by contributions from later processes such as the recombination of photocarriers and extrinsic processes at the interface with the electrode.

\subsection{Geometrical effects in injection current}
\begin{figure}
    \centering
    \includegraphics[width=0.8\linewidth]{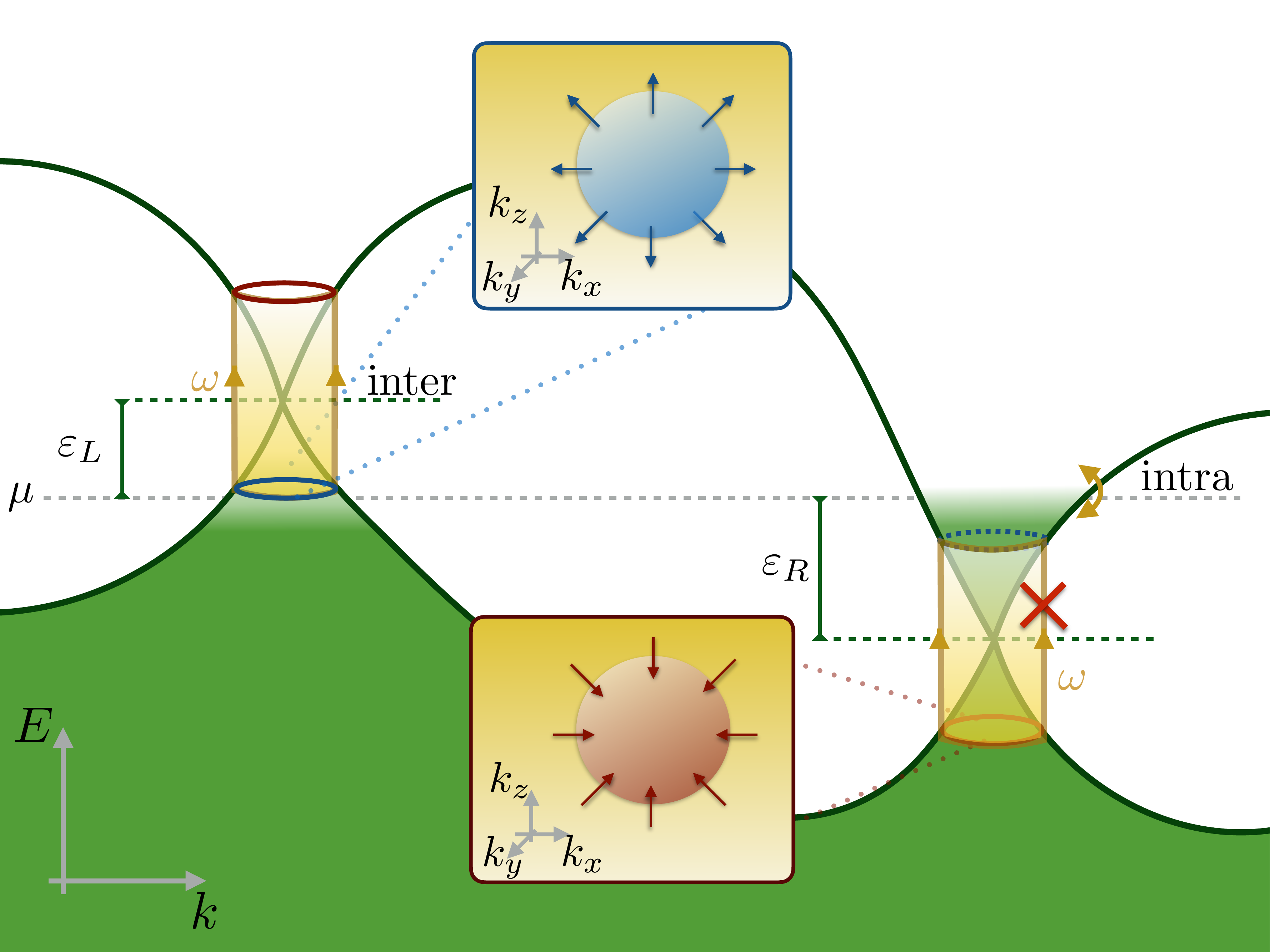}
    \caption{(Color Online)
    Quantized CPGE in Weyl semimetals. CPGE in Weyl semimetals shows a quantization corresponding to the Weyl charge.
    Since Weyl and anti-Weyl nodes contribute to the CPGE with opposite signs,
    the observation of the quantized CPGE requires that the Weyl and anti-Weyl points are not located at the same energy and optical transition takes place for only one node.
    Ref.~\citen{deJuan17}. \copyright[2017]{CC BY 4.0.}
    }
    \label{fig: qcpge}
\end{figure}

Injection current from the group velocities of the photocarriers also has geometrical aspects. 
While the shift current is related to the Berry connection difference between the two bands associated with optical transition, the injection current contribution for CPGE is governed by the Berry curvature in the case of two band systems.
This connection to the Berry curvature originates from the difference in the optical transition rates for left and right circularly polarized lights, which is proportional to $\mathrm{Im}[v_{x,12} v_{y,12}]/\epsilon_{12}^2$ and reduces to the Berry curvature in the two band cases.
One interesting consequence is the quantization of CPGE in Weyl semimetals.\cite{deJuan17}
In Weyl semimetals, two bands cross at a single Weyl point with a linear dispersion around it and the Weyl point behaves as a source or a sink of Berry flux of a quantized amount.
The above connection of the CPGE to the Berry curvature and the existence of the Weyl point lead to the relationship,\cite{deJuan17}
\begin{align}
   \frac{1}{2}\left[\frac{dj_{\circlearrowright}}{dt}- \frac{dj_{\circlearrowleft}}{dt} \right]= \frac{4\pi \alpha e}{h} I C,
\end{align}
where $j_{\circlearrowright}$ and $j_{\circlearrowleft}$ are photocurrent under left and right circular polarized light, $\alpha$ is the fine structure constant, $I$ is the intensity of the light, and $C$ is the charge of the Weyl point.
Thus the part of photocurrent that grows linearly in time, i.e., injection current, shows a quantization into the Weyl charge.
Since Weyl and anti-Weyl nodes contribute to the CPGE with opposite signs,
the observation of the quantized CPGE requires Weyl and anti-Weyl points are not located at the same energy and contribution from only one node appears (Fig. \ref{fig: qcpge}).  
For example, the Weyl semimetal should lack any mirror symmetry because a mirror partner of Weyl and anti-Weyl nodes appear at the same energy in the presence of mirror symmetry. 

A candidate system to observe the quantized CPGE is a multifold chiral fermion that is an analog of the Weyl fermion with higher spin representations, where more than two bands cross at a single point with linear dispersions.~\cite{Flicker18,deJuan20}
For example, measurements of CPGE were performed in multifold chiral fermion materials, RhSi \cite{rees2020helicity} and CoSi \cite{ni2021giant}, which belong to the space group P2$_1$3 (\#198) and lack mirror symmetry.
A measurement of photocurrent in RhSi reported that the CPGE shows a plateau structure which is suggestive to be consistent with the quantization. However, such plateau structure may not solely originate from the Weyl structure because the frequency dependence of the relaxation time $\tau$ and contributions to the CPGE from other bands should be present.\cite{rees2020helicity}
Therefore a more direct observation of the quantized CPGE remains a future problem.



\section{Geometrical Phenomena in Diabatic Processes}

As we have seen in Sec.~\ref{sec: berry}, geometric notions naturally appear when the motion of electrons is restricted to a particular Bloch band, which is indeed the case when the external field is so weak that interband transitions are negligible. On the other hand, we have also seen in Sec.~\ref{sec: NLOR} that resonant excitation via the intense light field accompanies the change in the geometry, which contributes to nonlinear responses as exemplified in the shift current.
In this section, let us review the geometric phase appearing in the adiabatic time evolution and nonperturbative process described as a correction to the adiabatic limit.

\subsection{Adiabatic time evolution and geometric phase}
Let us consider the Bloch electrons described by 
\begin{equation}
\hat{H}(\bm{k})|u_{n,\bm{k}}\rangle=\epsilon_n(\bm{k})|u_{n,\bm{k}}\rangle
\end{equation}
and their dynamics under an external electric field.

In the velocity gauge, the electric field is encoded via the vector potential as $\hat{H}(\bm{k})\to\hat{H}(\bm{k}(t))$ with $\bm{k}(t)=\bm{k}-q\bm{A}(t)/\hbar$.
In the weak field limit, we expect that the dynamics of electrons is constrained on the Bloch band in the initial configuration. Namely, the solution of the time-dependent Schr\"odinger equation $i\hbar\partial_t|\Psi(t)\rangle=\hat{H}(\bm{k}(t))|\Psi(t)\rangle$ should be written as $|\Psi(t)\rangle\simeq|u_{n,\bm{k}(t)}\rangle e^{i\gamma_n(t)}$ with $\gamma$ being an appropriate phase factor.
To examine this expectation quantitatively, let us take the above ansatz as a basis and expand $|\Psi(t)\rangle$ as\cite{DeGrandi2010}
\begin{equation}
|\Psi(t)\rangle=\sum_n C_n(t)|u_{n,\bm{k}(t)}\rangle e^{i\gamma_n(t)}.
\end{equation}
Then the time-dependent Schr\"odinger equation can be rewritten as
\begin{equation}
\dot{C}_n(t) 
=  \frac{iq}{\hbar}\sum_{m\neq n} \bm{E}(t)\cdot
\bm{\xi}_{nm}(\bm{k}(t)) e^{i(\gamma_m(t)-\gamma_n(t))}C_m(t)
\end{equation}
with $\bm{\xi}_{nm}(\bm{k})=\langle u_{n,\bm{k}}|i\partial_{\bm{k}}|u_{m,\bm{k}}\rangle$.
Here we set the phase factor as $\gamma=\gamma^{\text{d}}+\gamma^{\text{B}}$ to eliminate $C_n$ term in the left hand side,
where $\gamma^{\text{d}}$ and $\gamma^{\text{B}}$ are the so-called dynamical phase and the Berry phase, respectively, defined as 
\begin{equation}
\gamma_n^\text{d}(t)=-\frac{1}{\hbar}\int_{t_0}^t dt^\prime\epsilon_n(\bm{k}(t^\prime)),
\quad\gamma_n^\text{B}(t) 
= \int_{\bm{k}(t_0)}^{\bm{k}(t)} d\bm{k}\cdot\bm{a}_{n}(\bm{k}).
\end{equation}
The appearance of the Berry phase factor is a natural consequence of a constrained dynamics on a particular Bloch band.

The above equation can be formally solved in terms of the time-ordered exponential, as in the standard time-dependent perturbation theory in the interaction picture.
If we truncate the series at the first order, the approximate solution is obtained as
\begin{equation}
C_n(t) 
\simeq C_n(t_0)+ \sum_{m\neq n} W_{nm}(t)C_m(t_0).
\end{equation}
with
\begin{equation}
W_{nm}(t) = \frac{iq}{\hbar}\int_{t_0}^{t}dt \bm{E}(t)\cdot\bm{\xi}_{nm}(\bm{k}(t)) e^{i(\gamma_m(t)-\gamma_n(t))}.\label{eq:interband-amp}
\end{equation}

The interband transition amplitude $W_{nm}$ can be expanded into a power series of $\bm{E}$ using the relation $e^{i(\gamma^\text{d}_m(t)-\gamma^\text{d}_n(t))}=i\hbar(\epsilon_m(\bm{k}(t))-\epsilon_n(\bm{k}(t)))^{-1}\partial_t e^{i(\gamma^\text{d}_m(t)-\gamma^\text{d}_n(t))}$ and performing partial integration recursively. If we neglect $\dot{\bm{E}}(t)$, the leading order expression is obtained as
\begin{align}
W_{nm}(t) &\simeq  \frac{q\bm{E}(t)\cdot\bm{\xi}_{nm}(\bm{k}(t))}{\epsilon_n(\bm{k}(t))-\epsilon_m(\bm{k}(t))}e^{i(\gamma_m(t)-\gamma_n(t))},\label{eq:interband-amp-linear}
\end{align}
where we have assumed that the external field is absent at $t=t_0$.
This correction linear in the field amplitude contributes to the Hall current as 
\begin{align}
&\frac{1}{\hbar}\langle\Psi(t)|\partial_{\bm{k}}\hat{H}(\bm{k}(t))|\Psi(t)\rangle\nonumber\\&=
\sum_{ml}\frac{C_m^\ast(t) C_l(t)}{\hbar}e^{i(\gamma_l(t)-\gamma_m(t))}[
\delta_{ml}\partial_{\bm{k}}\epsilon_m+i(\epsilon_m-\epsilon_l)
\bm{\xi}_{ml}]\nonumber\\&\simeq
\frac{1}{\hbar}\partial_{\bm{k}}\epsilon_n(\bm{k}(t))
+2\text{Im}
\sum_{m\neq n}\frac{W_{nm}(t)}{\hbar}e^{i(\gamma_n(t)-\gamma_m(t))}(\epsilon_m-\epsilon_n)
\bm{\xi}_{mn}\nonumber\\&\simeq
\frac{1}{\hbar}\partial_{\bm{k}}\epsilon_n(\bm{k}(t))
-\frac{2q}{\hbar}\text{Im}
\sum_{m\neq n}\bm{E}(t)\cdot\bm{\xi}_{nm}\bm{\xi}_{mn}
\nonumber\\&=
\frac{1}{\hbar}\partial_{\bm{k}}\epsilon_n(\bm{k}(t))-
\frac{q}{\hbar}\bm{E}(t)\times\bm{F}_{n}(\bm{k}(t)),
\end{align}
assuming that the electron is initially at $n$th band, i.e., $C_n(t_0)=1$ and $C_{m\neq n}(t_0)=0$.
Here we have used $\text{Im}\sum_{m\neq n} \bm{E}\cdot\bm{\xi}_{nm}\bm{\xi}_{mn}=
 \bm{E}\times (\partial_{\bm{k}}\times \bm{a}_{n})/2$.

\subsection{Nonreciprocal Landau-Zener tunneling}
The interband transition amplitude represented by Eq.~(\ref{eq:interband-amp}) includes nonperturbative effects as well as the perturbative contribution described above. To see this, it is convenient to change the variable of integral from $t$ to $\bm{k}$, which leads to 
\begin{equation}
W_{nm}(t) = i\int_{k(t_0)}^{k(t)}dk\,\xi_{nm}(k) 
e^{
i\int_{k(t_0)}^{k} dk^\prime
(-\frac{\epsilon_m-\epsilon_n}{q E}
+ a_{m}-a_{n})}.
\end{equation}
Here for simplicity we focus on one-dimensional cases with a dc electric field.
This expression implies that the suppression of the interband transition originates from the rapid oscillation of the integrand leading to cancellation, and nonperturbative treatment is necessary for precise evaluation.
We need to interpret the above integral as a contour integral on the complexified $k$ plane\cite{Dykhne1962,Davis1976,LandauLifshitz}, and apply the saddle point method.\cite{Fukushima2020,Kitamura2020-2}
Saddle points of the integrand are given by the condition of a vanishing (logarithmic) derivative, i.e.,  $-i\partial_k\ln\xi_{nm}(k)-(\epsilon_m-\epsilon_n)/qE+a_m-a_n=0$,
which implies that the saddle points appear in the vicinity of gap closing points $\epsilon_m=\epsilon_n$, and play a crucial role when the electron passes through anticrossing points (real $k$ points closest to complex gap closing points).

\begin{figure}
    \centering
    \includegraphics[width=\linewidth]{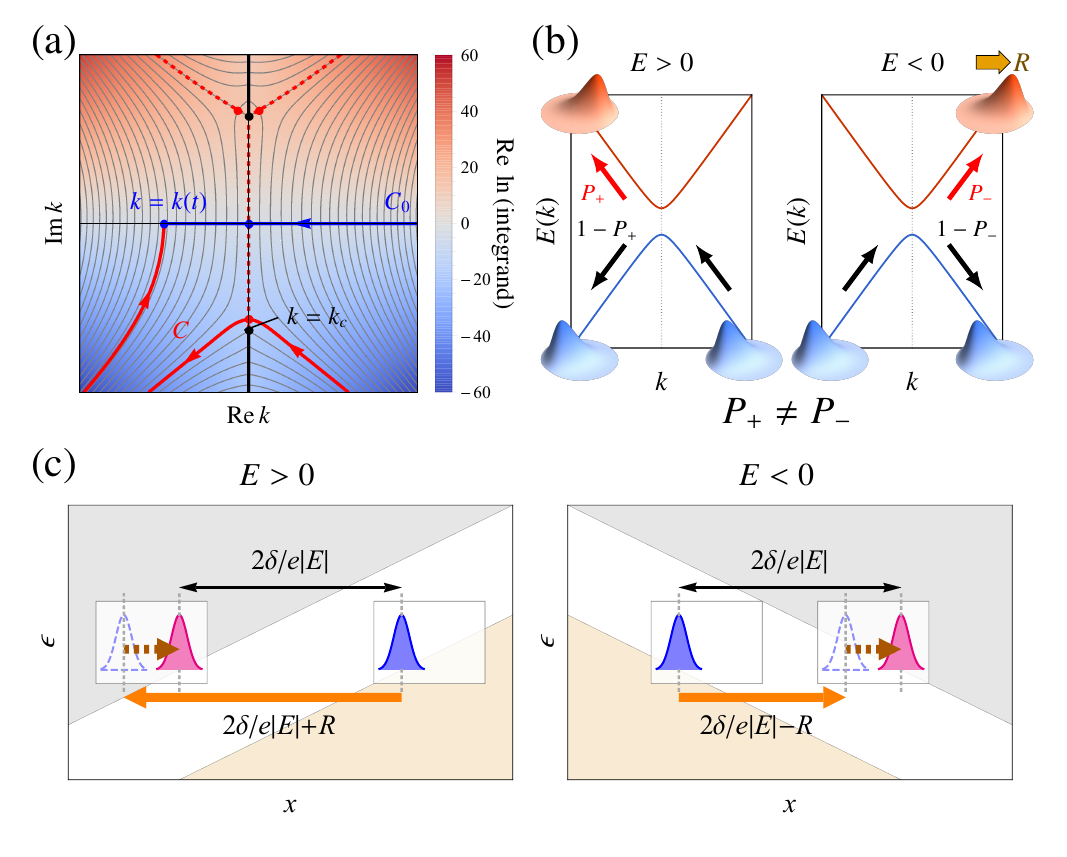}
    \caption{(Color Online)
    (a) Integrand of Eq.~(\ref{eq:wnm}) in the complexified momentum plane for a typical two-band system.
    Gray lines are the steepest descents, along which the imaginary part is constant. The original path of integration, labeled by $C_0$ here, is indicated by the blue line, which can be deformed into the solid red curve $C$ to apply the saddle point method.
    Saddle points indicated by red dots appear in the vicinity of gap closing points $k_c$ (shown as black dots) for small electric fields. The black thick lines extend from the gap closing point are the branch cut.
    Taken from Ref.~\citen{Kitamura2020-2}. \copyright[2020]{American Physical Society.} 
    (b) Schematic picture of nonreciprocal Landau-Zener tunneling. The tunneling probability for positive and negative electric fields $E$ is denoted as $P_{\pm}$, which becomes different in noncentrosymmetric crystals due to the underlying geometry of Bloch bands. The shift vector $R$ of a geometric origin is schematically depicted as a distortion of the wave packet for upper and lower bands.
    (c) The real space picture of nonreciprocal tunneling. The tunneling barrier effectively changes according to the shift vector, whose asymmetry leads to nonreciprocity. 
    Taken from Ref.~\citen{Kitamura2020}. \copyright[2020]{CC BY 4.0.}
    }
    \label{fig:tunnel}
\end{figure}

We note that the integral above involves the Berry phase factor, which implies that the geometrical effect should also appear in nonperturbative phenomena. 
On the other hand, the above expression is not suitable for exploring the geometrical effect, because the Berry phase here is defined for open paths in $k$ space and 
thus not a gauge invariant. It can be rewritten in terms of the shift vector, $R_{mn}=\partial_k \arg \xi_{nm}+a_m-a_n$, as\cite{Kitamura2020}
\begin{equation}
W_{nm}(t) = ie^{i\arg \xi_{nm}(k_0)}\int_{k(t_0)}^{k(t)}dk\,|\xi_{nm}(k)| 
e^{
i\int_{k(t_0)}^{k} dk^\prime
(-\frac{\epsilon_m-\epsilon_n}{q E}
+ R_{mn})},\label{eq:wnm}
\end{equation}
with which we can perform analytic continuation of the integrand regardless of gauge choices.
We show a contour plot of the integrand for a typical two-band problem with an anticrossing in Fig.~\ref{fig:tunnel}(a). The original path of the integral $k\in[k(t_0),k(t)]$ lies on the real axis, which can be deformed into the segments along the steepest descents. The perturbative expression Eq.~(\ref{eq:interband-amp-linear}) corresponds to the contribution from the path around the end point $k=k(t)$, while the nonperturbative correction should appear due to the contribution from the saddle point, whose asymptotic form in $E$ is given as
\begin{equation}
|W_{nm}|^2 \simeq \exp\left[
-2\text{Im}\int_{k(t_0)}^{k_c} dk
\left(-\frac{\epsilon_m-\epsilon_n}{q E}
+ R_{mn}\right)\right],
\end{equation}
where $k_c$ is the gap closing point (that gives a negative exponent).
This expression with an essential singularity cannot be captured in a perturbative treatment, and this nonperturbative transition is known as the Landau-Zener tunneling. In noncentrosymmetric crystals with a nonzero shift vector, we have a geometric prefactor shown above.\cite{Berry1990,Joye1991,Wu2008,Kitamura2020,Takayoshi2021} This prefactor depends on the sign of the electric field, and results in nonreciprocal behavior (rectification effect): When the electric field is inverted, we need to replace $k_c$ by $k_c^\ast$ to have a negative exponent, which then inverts the sign of the geometric exponent.
The ratio between the tunneling probability $P=|W_{nm}|^2$ for positive and negative electric fields is given as $P(+|E|)/P(-|E|)=\exp[
-2\text{Im}\int_{k_c^\ast}^{k_c} dk
R_{mn}]$, which does not depend on the field strength $|E|$ reflecting its geometric nature.
This inequivalence of the tunneling probability for left and right propagation (i.e., nonreciprocal Landau-Zener tunneling) is schematically depicted in Fig.~\ref{fig:tunnel}(b). 
We also show a real-space picture with different effective barrier thicknesses in Fig.~\ref{fig:tunnel}(c).

Due to this nonreciprocity in the tunneling process, the resultant carrier density differs in positive and negative electric fields, which leads to a nonreciprocal current response.\cite{Kitamura2020-2} 
Here the nonreciprocal current response mainly comes from contributions of the group velocity with asymmetric distribution of tunneling electrons.
Note that this mechanism contrasts with the case of the shift current in the optical regime, although it shares the same geometric quantity as an origin. 
Specifically, in the case of the shift current, the distribution function remains symmetric but the generated electric polarization upon the creation of an electron-hole pair contributes to the nonreciprocal transport.

\section{Discussions}
We have reviewed several nonlinear and nonequilibrium phenomena in which the notion of geometry and topology becomes important. Floquet engineering of quantum materials has a large potential for controlling quantum phases of matter by tailoring external light fields. In particular, its application to strongly correlated systems such as magnets and superconductors is still developing, aiming at creating and controlling exotic quantum phases. Also, the BPVE in inversion broken materials including shift current can be a novel probe for geometric properties of quantum materials given its close connection to the geometry of Bloch electrons, and has a potential future application to photoelectric conversion such as efficient solar cell action and light detector. In closing this review paper, we discuss some of the possible future directions in these research fields.

For the Floquet engineering of geometrical responses, an interesting future direction is the use of  field profiles peculiar to multiple laser lights. 
In this paper, we have mainly reviewed engineering using monochromatic circularly-polarized light. 
The key point there is that the Floquet effective Hamiltonian inherits the symmetry of the field profile,
with which unconventional interactions with broken time-reversal symmetry emerge and induce geometric responses in the case of circularly-polarized lights.
The bicircular light, which consists of a pair of left and right circularly-polarized lights with commensurate frequencies, provides 
a new platform for Floquet engineering with new types of symmetry breaking, since its field profile has a discrete rotational symmetry.~\cite{Nag2019,Ikeda2021,Trevisan2022}
A pair of laser lights with incommensurate frequencies leads to quasiperiodic time dependence, 
which has also various potential applications. \cite{Martin2017,Verdeny2016,Ray2019,Else2020-2}
Quasiperiodic structures are known to be regarded as a projection of a periodic structure in higher dimensions, and indeed driving 
with two incommensurate frequencies can be analyzed in the Sambe space with two Floquet indices. 
Recently, the physics of spatial quasiperiodicity is intensively explored in stacked 2D van der Waals materials with a twist angle. \cite{Andrei2020} 
It is interesting to explore its dynamical realization.

There are many issues in nonlinear 
transport phenomena still to be explored. 
Conventional Boltzmann transport theory captures
the leading order terms in $\tau$, while the 
sub-leading order terms contain the information 
on quantum geometry of the electronic states \cite{PhysRevB.106.125114}
Another direction is to consider the nonlinear 
transport in the dissipationless currents. Two 
representative examples are the superconducting 
current and quantum Hall current. In the former,  
extensive theoretical and experimental studies have 
been done recently on the superconducting diode 
effect where the critical current to break the 
superconductivity depends on its direction \cite{Ando2020,He_2022,doi:10.1073/pnas.2119548119,PhysRevLett.128.037001}.
In the latter case, the edge channel picture 
predicts the linear relation between the Hall current $I$ and the chemical potential 
difference $\Delta \mu$ between the two edges, 
which defines the quantized Hall conductance.
Namely, there should be no nonlinear transport.
However, this can be modified when the 
electron-electron interaction is taken into account,
which requires further studies.

Shift current in correlated electron systems will be an interesting research topic. In this paper, we reviewed shift current responses of magnons in multiferroic magnets and phonons in electron-phonon coupled systems. There exist many other bound states and elementary excitations accompanying electric polarization in correlated electron systems with broken inversion symmetry. Their optical excitations may support a novel photoelectric conversion functionality and will be an interesting future problem. 

While we primarily focused on shift current in this review, the ballistic current is ubiquitously present in the BPVE in inversion broken materials due to the asymmetric scattering of photoexcited electrons and gives an important contribution. In particular, ballistic current contains a contribution proportional to the relaxation time $\tau$ and can be large when $\tau$ is long, while shift current does not depend on $\tau$. The analysis of ballistic current is difficult because it involves complicated scattering processes of photoexcited electrons. While there exist efforts to quantitatively study ballistic current \cite{Dai21-prl,Dai21-prb}, more theoretical development for studying ballistic current is awaited.

Finally, nonlinear responses in nonequilibrium systems are also an interesting direction. Nonequilibrium states can potentially offer novel nonlinear responses that cannot be achieved in the equilibrium states that have been conventionally studied. For example, the exciton-polariton is a quasiparticle realized in the nonequilibrium state in a semiconductor coupled to a photon cavity and is made of the hybridization of excitons in the semiconductor and cavity photons. With noncentrosymmetric crystals, the exciton-polaritons can carry photocurrent with the shift current mechanism.\cite{Morimoto20-polariton} Studying the role of dissipation in the BPVE will be an interesting topic.

\bibliographystyle{jpsj}
\bibliography{reference,footnotes}

\begin{thebibliography}{100}
\bibitem{footnotesk1}
In the interaction picture, the photon operator becomes time-dependent as
  $\hat{a}(t)=e^{i\Omega \hat{a}^\dagger\hat{a}t}\hat{a}e^{-i\Omega
  \hat{a}^\dagger\hat{a}t}=\hat{a}e^{-i\Omega t}$, which implies that
  $\hat{H}_{+1}$ eliminates one photon from the external photon field.
\bibitem{footnotesk2}
The magnetic flux penetrating the triangle enclosed by sites $ijk$ in terms of
  the Peierl phase is given as $\arg[t_{ij}t_{jk}t_{ki}]=\oint_{i\to j\to k\to
  i}d\bm{R}\cdot\bm{A}(\bm{R})$.
\bibitem{footnotetm1}
Note that the Berry connection is defined with an extra minus sign in
  Ref.~[\citen{Morimoto-Nagaosa16}].
\end{thebibliography}

\profile{Takahiro Morimoto}{was born in Okayama, Japan in 1985. He obtained PhD at the University of Tokyo in 2012. He was a postdoctoral researcher at RIKEN and a Moore foundation postdoctoral fellow at University of California, Berkeley. He is an associate professor at the University of Tokyo since 2019.
His research interest is in theoretical condensed matter physics, with an emphasis on topological phases of matter and their response phenomena.}

\profile{Sota Kitamura}{was born in 1990 in Aomori prefecture, Japan. He received his PhD from the University of Tokyo in 2017. 
He has worked as a postdoctoral researcher at MPI-PKS, Dresden, from 2017 to 2019. Currently, He is a research associate at the University of Tokyo. His research focuses on theoretical condensed-matter physics, particularly nonequilibrium and nonperturbative phenomena in solids.}

\profile{Naoto Nagaosa}{
was born in Hyogo Prefecture in 1958, and graduated from Department of Applied physics, The University of Tokyo in 1980. From 1983 to 1986, he was a research associate in Institute for Solid State Physics, Univ. Tokyo, and received a D.Sci from Univ. Tokyo in 1986. From 1988 to 1990, he worked as a visiting scientist at Department of Physics, Massachusetts Institute of Technology, before joining the Department of Applied Physics in Univ. Tokyo where he is now a professor. From 2013 he has joint appointment with the Deputy Director of the RIKEN Center for Emergent Matter Science (CEMS). His research field is theoretical condensed-matter physics, especially involving the strong electron correlation, optical responses of solids, topological aspects of condensed matter, and superconductivity. }

\end{document}